\documentclass[aps, twocolumn,showpacs,superscriptaddress,floatfix]{revtex4-1}
\usepackage[paperwidth=8.5in,paperheight=11in,centering,hmargin=0.75in,vmargin=0.9in]{geometry}
\usepackage{graphicx}
\usepackage{hyperref}
\usepackage{amsmath}

\makeatletter
\let\cat@comma@active\@empty
\makeatother

\newcommand{\nocontentsline}[3]{}
\newcommand{\tocless}[2]{\bgroup\let\addcontentsline=\nocontentsline#1{#2}\egroup}

\begin{document}

\title{String patterns in the doped Hubbard model} 
\date{\today}
\author{Christie S. Chiu}
\author{Geoffrey Ji}
\affiliation{Department of Physics, Harvard University, Cambridge, Massachusetts 02138, USA}
\author{Annabelle Bohrdt}
\affiliation{Department of Physics and Institute for Advanced Study, Technical University of Munich, 85748 Garching, Germany}
\affiliation{Department of Physics, Harvard University, Cambridge, Massachusetts 02138, USA}
\affiliation{Munich Center for Quantum Science and Technology (MCQST), Schellingstr. 4, D-80799 M\"unchen, Germany}
\author{Muqing Xu}
\affiliation{Department of Physics, Harvard University, Cambridge, Massachusetts 02138, USA}
\author{Michael Knap}
\affiliation{Department of Physics and Institute for Advanced Study, Technical University of Munich, 85748 Garching, Germany}
\affiliation{Munich Center for Quantum Science and Technology (MCQST), Schellingstr. 4, D-80799 M\"unchen, Germany}
\author{Eugene Demler}
\affiliation{Department of Physics, Harvard University, Cambridge, Massachusetts 02138, USA}
\author{Fabian Grusdt}
\affiliation{Department of Physics, Harvard University, Cambridge, Massachusetts 02138, USA}
\affiliation{Munich Center for Quantum Science and Technology (MCQST), Schellingstr. 4, D-80799 M\"unchen, Germany}
\author{Markus Greiner}
\email[Corresponding author. Email: ]{greiner@physics.harvard.edu.}
\author{Daniel Greif}
\affiliation{Department of Physics, Harvard University, Cambridge, Massachusetts 02138, USA}

\begin{abstract}
	Understanding strongly correlated quantum many-body states is one of the most difficult challenges in modern physics.
	For example, there remain fundamental open questions on the phase diagram of the Hubbard model, which describes strongly correlated electrons in solids.
	In this work we realize the Hubbard Hamiltonian and search for specific patterns within the individual images of many realizations of strongly correlated ultracold fermions in an optical lattice.
	Upon doping a cold-atom antiferromagnet we find consistency with geometric strings, entities that may explain the relationship between hole motion and spin order, in both pattern-based and conventional observables.
	Our results demonstrate the potential for pattern recognition to provide key insights into cold-atom quantum many-body systems.
\end{abstract}


\maketitle

Quantum superposition describes quantum systems as simultaneously realizing different configurations.
Such behavior is believed to be at the heart of phenomena in strongly correlated quantum many-body systems, which cannot be described by single-particle or mean-field theories.
An intriguing consequence of the superposition principle is the existence of hidden order in correlated quantum systems: although every individual configuration is characterized by a particular pattern, the average over these configurations leads to an apparent loss of order.
By contrast, instantaneous projective measurements have the potential to reveal these underlying patterns.

One notable example of a system with hidden order is the one-dimensional (1D) Fermi-Hubbard model at strong coupling \cite{Lieb1968, Woynarovich1982}.
Although 1D chains with additional holes or particles beyond an average of one particle per site (doped) yield average two-point spin correlations which decay more rapidly with distance than chains with an average of one particle per site (half-filled), this magnetic ordering can be revealed by accounting for the fluctuating positions across individual configurations of the additional dopants within each chain.
The apparent loss of magnetic order is in fact hidden order, hidden by the dopants and their varying positions \cite{Ogata1990, Kruis2004}.
Although direct detection of this hidden string order remains inaccessible in solids, experiments with ultracold atoms enable projective measurements, or ``snapshots", and generally can provide access to such structures \cite{Endres2011}.
In particular, quantum gas microscopy \cite{Gross2017} enables site-resolved imaging and access to correlators which have been constructed to reveal the hidden order \cite{Hilker2017}.

The hidden order in 1D is well understood, but the physics of the 2D Hubbard model is fundamentally more complex due to an intricate interplay between spin and charge degrees of freedom; as a result, formulating an appropriate correlation function to search for hidden order becomes considerably more challenging.
The 2D Hubbard model is believed to capture the rich physics of high-temperature superconductivity and other phases \cite{Emery1987,Lee2006,Keimer2015} such as the strange metal, stripe, antiferromagnet (AFM), or pseudogap phase, but a unified understanding of these phenomena is still lacking.
For example, the behavior of individual dopants in an AFM is not agreed upon, including whether hidden string order is present and dopants hide magnetic correlations by shifting the positions of a string of spins.
Quantum gas microscopy, however, provides a perspective that goes beyond the framework of two- or multi-point correlations.
Hidden string order can be searched for directly within individual snapshots of the quantum mechanical wavefunction, where quantum fluctuations are resolved.

Here we perform a microscopic study of the hole-doped Fermi-Hubbard model and report indications of string patterns in 2D over a wide doping range.
Our measurements use ultracold fermions in an optical lattice down to the lowest currently achievable temperatures, where at low doping AFM correlations extend across the system size \cite{Mazurenko2017}.
We identify string patterns in individual projective measurements and compare them with predictions from microscopic theoretical approaches.

\tocless{\section*{Candidate theories for the doped Hubbard model}}

\begin{figure}
	\begin{center}
		\includegraphics[width=\columnwidth]{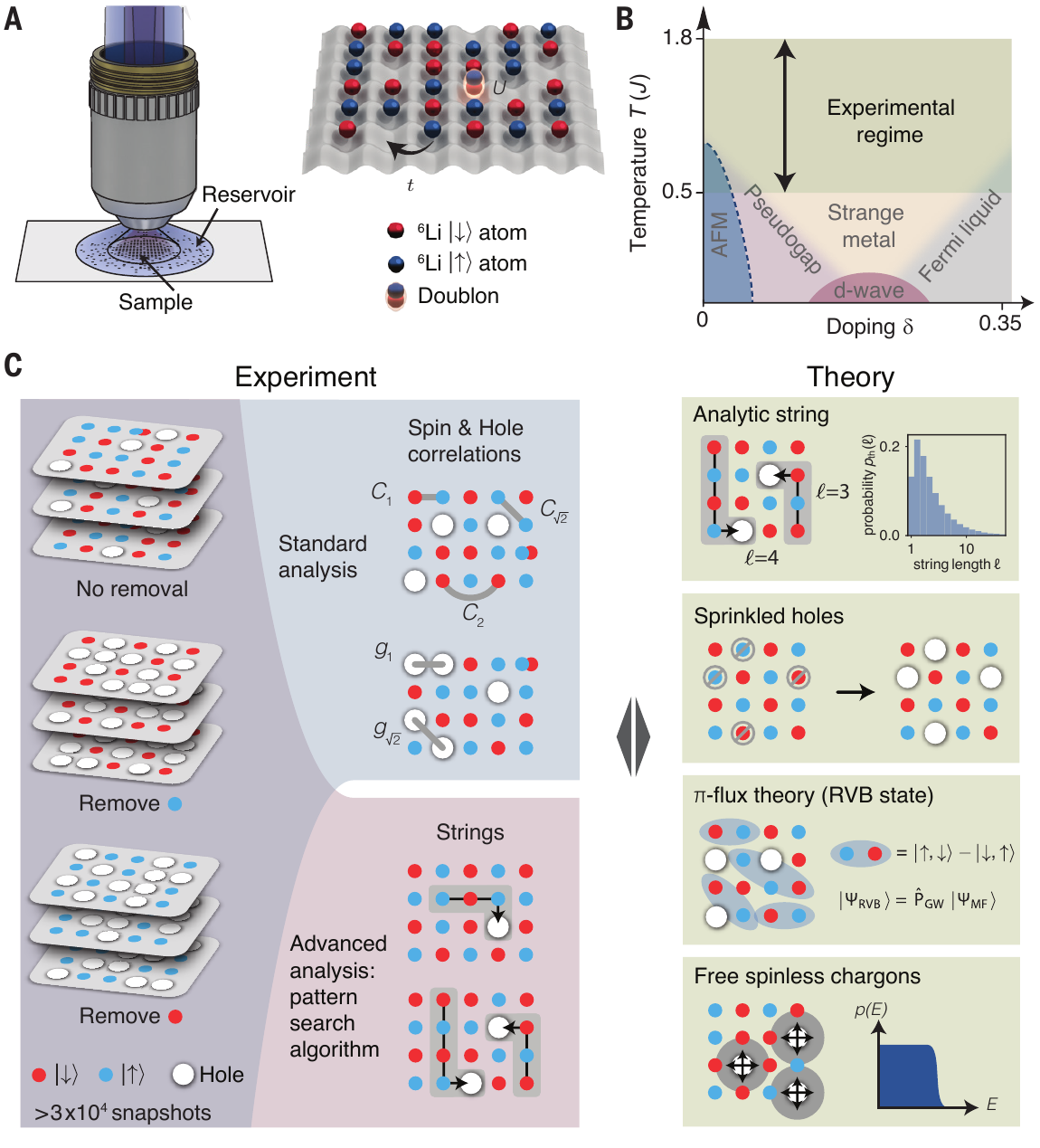}
		\label{fig1:intro}
	\end{center}
	\vspace{-25pt}
	\caption{\textbf{Quantum simulation of the Hubbard model.} \textbf{(A)} Quantum gases trapped in optical lattices realize the Hubbard model with tunable on-site interaction $U$ and nearest-neighbor hopping $t$. Quantum gas microscopy enables site-resolved readout of the quantum state. \textbf{(B)} Schematic of the conjectured phase diagram of the finite-size 2D Hubbard model with the experimentally accessed regime (green shading). \textbf{(C)} Outline of experimental observables used and theoretical models evaluated. We evaluate theories using both standard observables and pattern-recognition-based observables using snapshots of the quantum state.}
	\vspace{-10pt}
\end{figure}

We study the Fermi-Hubbard model, which is defined by the Hamiltonian
\begin{equation}
\hat{\mathcal{H}} = -t \sum_{\sigma = \uparrow, \downarrow} \sum_{\langle \mathbf{i}, \mathbf{j} \rangle} \left( \hat{c}^\dagger_{\mathbf{i},\sigma} \hat{c}_{\mathbf{j},\sigma} + {\rm h.c.} \right)+ U \sum_{\mathbf{j}} \hat{c}^\dagger_{\mathbf{j},\uparrow} \hat{c}_{\mathbf{j},\uparrow} \hat{c}^\dagger_{\mathbf{j},\downarrow} \hat{c}_{\mathbf{j},\downarrow}
\end{equation}
(see Fig.~1A).
The first term describes tunneling of amplitude $t$ of spin-$1/2$ fermions $\hat{c}_{\mathbf{j},\sigma}$ with spin $\sigma$ between adjacent sites $\mathbf{i}$ and $\mathbf{j}$ of a two-dimensional square lattice.
The second term includes on-site interactions of strength $U$ between fermions of opposite spin.
We consider the strongly correlated regime, where $U \gg t$ and doubly occupied sites are energetically costly.

The Fermi-Hubbard model is well understood when the band is half filled at an average of one particle per site (Fig.~1B).
For temperatures $T \ll J$, where $J = 4 t^2 / U$ is the super-exchange coupling, AFM correlations appear.
Although these magnetic correlations are finite-ranged at non-zero temperatures, sufficiently cold finite-size systems can have AFM order across the entire system \cite{Mazurenko2017}.

Much less is known about the doped Fermi-Hubbard model.
However, it is understood that dopant delocalization for kinetic energy minimization competes with spin interactions in the background AFM.
Experiments on the cuprates have also shown that at temperatures $T < J$ and between $10$ and $20\%$ doping, the pseudogap phase crosses over to the strange metal, located above the superconducting dome \cite{Lee2006}.
The two novel metallic phases (pseudogap and strange metal) defy a description in terms of conventional quasiparticles and still lack a unified theoretical understanding.

Although phenomenological, numeric, and mean-field (MF) approaches have provided key insights in the past, quantum gas microscopy is naturally suited to assess microscopic theoretical approaches.
One such theory is Anderson's resonating valence bond (RVB) picture \cite{Anderson1987}, which considers trial wavefunctions of free holes moving through a spin liquid comprised of singlet coverings.
We consider one particular class of RVB wavefunctions which have been studied extensively, called $\pi$-flux states \cite{Wen1996}.
They stem from a mean-field density matrix $\hat{\rho} = \hat{\mathcal{P}}_{\rm GW} e^{- \hat{\mathcal{H}}_{\rm MF} / k_{\rm B} T} \hat{\mathcal{P}}_{\rm GW}$, where $k_{\rm B}$ is Boltzmann's constant, $\hat{\mathcal{P}}_{\rm GW}$ is the Gutzwiller projection, and $\hat{\mathcal{H}}_{\rm MF}$ is the quadratic Hamiltonian of itinerant fermions on a square lattice with a Peierls phase of $\pi$ per plaquette (see section 6.1 of \cite{SI} for details).
Snapshots of the trial state in the Fock basis can be obtained by Monte-Carlo sampling, with temperature $T$ as a free fit parameter \cite{Gros1989}.

A second microscopic approach that we examine is the geometric-string theory \cite{Grusdt2018}, where AFM order at half-filling is hidden in doped states via hole motion.
This theory extends earlier work \cite{Bulaevskii1968, Brinkman1970, Beran1996} and establishes a relationship between the AFM parent state at half filling and the strongly correlated quantum states at finite doping.
Here, holes move through the parent AFM by displacing each spin along its trajectory by one lattice site, while the AFM quantum state remains otherwise unmodified; this is the frozen-spin approximation \cite{Grusdt2018a}.
The delocalization of each hole can then be described as a superposition state of hole trajectories, or geometric strings, whose lengths $\ell$ depend on the strength of AFM correlations and the ratio of the kinetic energy $t$ to the super-exchange $J$.
For any given temperature, a distribution function $p_{\rm th}(\ell)$ of string lengths can be obtained by sampling a Boltzmann distribution of string states (Fig. 1C).

We directly assess these microscopic theoretical approaches with a quantum gas microscope, which provides projective measurements of the quantum mechanical wavefunction for the doped Hubbard model in the parity-projected Fock basis.
Our experimental setup consists of a balanced two-component gas of fermionic Lithium in the lowest band of a square optical lattice \cite{Parsons2015}, with $U/t$ set to $8.1(2)$.
We selectively image one of the spin states or the total atom distribution \cite{Parsons2016}.
Entropy redistribution with a digital micro-mirror device enables a disk-shaped homogeneous system of approximately 80 sites with temperatures as low as $T/J = 0.50(4)$ \cite{Mazurenko2017}.
We alter the local chemical potential to dope the system, maintaining independent temperature control (section 7.1 of \cite{SI}).
We determine the doping from the single-particle occupation density and temperature from the nearest-neighbor spin correlator, both by comparing to numerics (section 2 of \cite{SI}).

\tocless{\section*{Pattern recognition of geometric strings}}

\begin{figure}
	\begin{center}
		\includegraphics[width=\columnwidth]{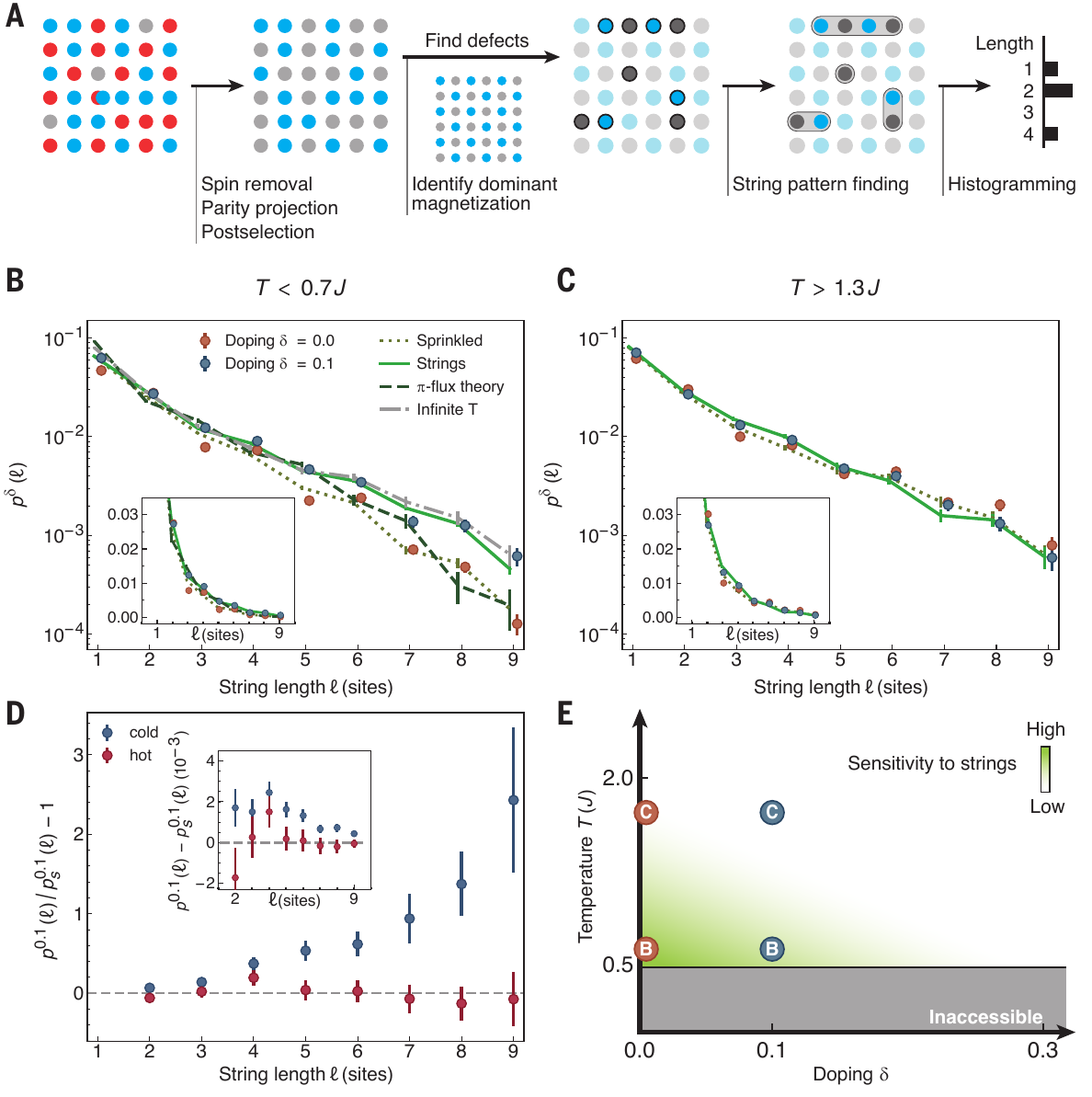}
		\label{fig2:hist}
	\end{center}
	\vspace{-25pt}
	\caption{\textbf{Measurement of string-pattern length histograms from site-resolved snapshots.} \textbf{(A)} Schematic explanation of the string-pattern identification algorithm (see text). \textbf{(B} and \textbf{C)} Change in string-pattern length histograms upon doping to $10\%$ and comparison with simulated models at $10\%$ doping, for temperatures below (B) and above (C) the superexchange energy $J$. The observable is only sensitive to doping in the colder dataset, and simulated strings seem to fit the doped experimental result best. Points have been slightly offset horizontally for readability and insets plot the same data on linear-linear axes. \textbf{(D)} Relative and absolute (inset) difference between doped and undoped pattern-length histograms, highlighting temperature-dependent sensitivity. The sprinkled-hole result is used for the undoped case as it accounts for the change in density. \textbf{(E)} Regions of the phase diagram examined in (B) and (C). The string-pattern observable has sensitivity at temperatures below $J$ and below intermediate doping. In (B), (C), and (D), histograms are normalized by the number of lattice sites analyzed and error bars represent 1 SEM from more than 5500 (half-filling, cold), 3500 (doped, cold), 2900 (half-filling, hot), and 4600 (doped, hot) images.}
	\vspace{-10pt}
\end{figure}

We design a pattern recognition algorithm for geometric strings that we apply to real-space snapshots where doublons and one of the two spin states have been removed (Fig.~2A).
Because geometric strings describe a relationship between doped and half-filled AFMs, we search for string-like patterns in the deviation between snapshots of the doped Hubbard model and an approximation to the AFM, an exact checkerboard.
For each image, we take the set of sites which deviate and extract string patterns using the following rules: (1) every string pattern is a connected subset of sites forming a path without branching points, (2) each site can be part of only one string pattern, (3) longer string patterns are favored, and (4) every string pattern must have at one end a site which is detected as empty, and therefore consistent with having a hole on that site.
We discuss alternate algorithms in section 3.4 of \cite{SI}.

We find that this algorithm is indeed sensitive to hole doping.
Figure 2B shows string-pattern length distributions $p^\delta(\ell)$ over pattern lengths $\ell$, averaged over experimental data at temperatures between $0.50(4)J$ and $0.70(3)J$.
As the sample is doped from half-filling to a doping $\delta$ of $10.0(8)\%$, the number of string patterns increases across the entire range of lengths.
The appreciable distribution of string patterns $p^0(\ell)$ detected at half-filling reflects the deviation of a quantum AFM from our checkerboard approximation and therefore should be considered as a baseline level.
This baseline can be reproduced through Heisenberg quantum Monte Carlo simulation (see section 3.3.3 of \cite{SI}), and is largely caused by the finite temperature and underlying SU(2) symmetry of the system.
We have lessened these contributions by reducing the analysis region to a diameter of 7 sites and post-selecting on the staggered magnetization.
In section 3.3 of \cite{SI}, we show that results are robust to the choice of postselection scheme and that the limited detection of one of the spin states causes only an overall factor decrease in string patterns detected.

Next we compare our experimental results to the simulation results of three microscopic models.
We make predictions by producing artificial images and evaluating them with our string pattern detection algorithm, such that the detection is common to experiment and theoretical simulation.
Beginning with the analytic string model, we generate images by randomly placing a number of holes into actual experimental images taken at half-filling, then randomly propagating each hole according to the analytically generated string length histogram (see Fig.~1C) and appropriately displacing the spins along the hole’s path.
Note that this approach preserves the SU(2) symmetry of the system.
The resulting string-pattern length distribution agrees with experimental data (see Fig.~2B for $10\%$ doping), even though the theory has no free parameters.

To verify whether our measured signal simply results from the introduction of holes rather than changes to the spin background, we next compare our experimental result with simulations where holes are artificially and randomly placed (``sprinkled") into experimental data taken at half-filling, equivalent to placing one-site-long strings.
The associated string-pattern length distribution $p_s^{\delta}(\ell)$ fails to explain the experimental results, revealing the nontrivial interplay of spin and charge degrees of freedom in the 2D doped Hubbard model.
Last, we compare our experimental result to $\pi$-flux states by fitting the nearest-neighbor spin correlator for an effective temperature and producing simulated images at $10\%$ doping, and find quantitative agreement with experiment at short pattern lengths, but a deficit at long lengths.

We repeat the measurements for a sample heated before lattice loading to investigate temperature effects.
Figure 2C shows experimental data at half-filling and at $10.1(8)\%$ doping, along with the simulated prediction, averaged over samples at temperatures between $1.3(1)J$ and $1.8(1)J$.
In contrast to colder temperatures, there is no statistically significant deviation between the experimental data with and without hole doping; $p^{0.1}(\ell) \approx p^0(\ell)$.
For these temperatures, spin ordering is so weak that the resulting string patterns may mask additional effects from doping.
These deviations appear to set an upper bound on the density of detectable string patterns (Fig.~2E); we therefore plot the pattern length distribution for high-temperature and half-filling as a reference for the cold temperature datasets in Fig.~2B (gray dash-dotted line).

\begin{figure}
	\begin{center}
		\includegraphics[width=\columnwidth]{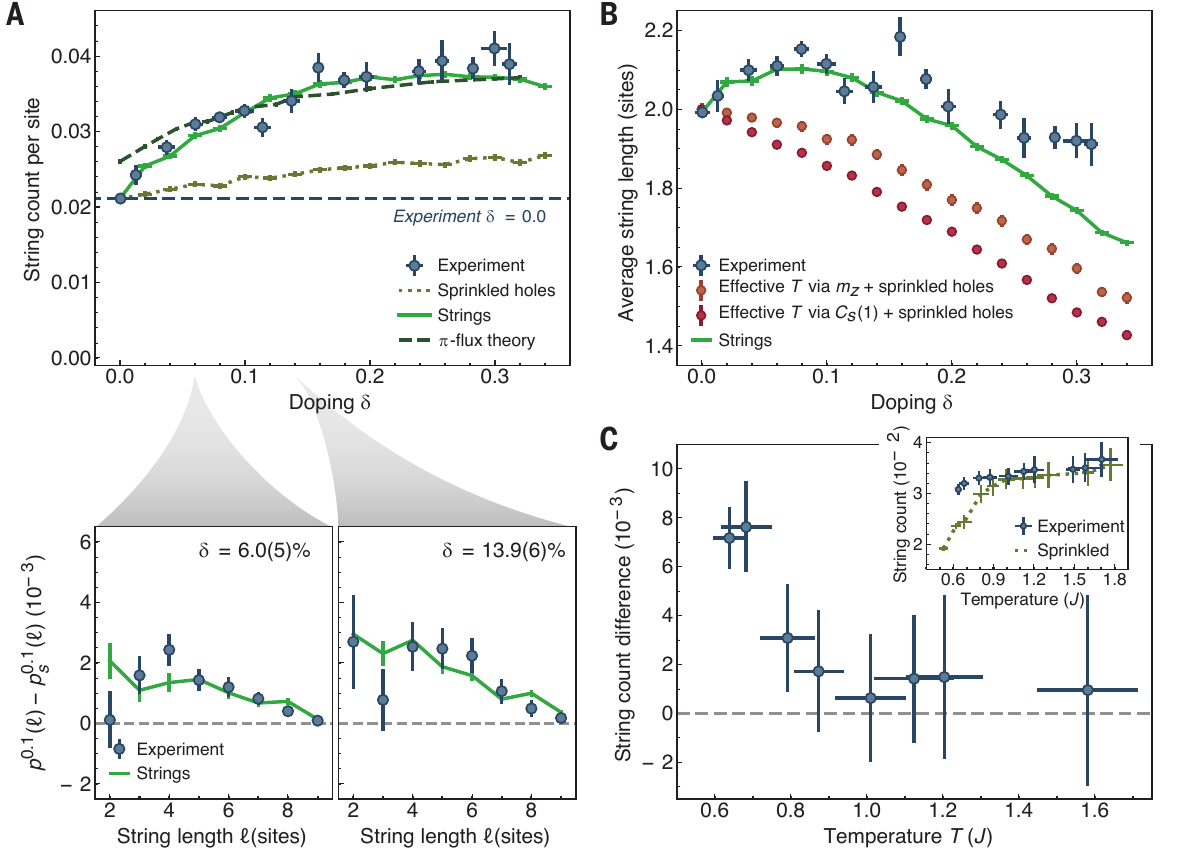}
		\label{fig3:sum}
	\end{center}
	\vspace{-25pt}
	\caption{\textbf{Detailed examination of the detected string patterns upon doping. (A)} (Top) Total number of string patterns exceeding length 2, normalized by the system size, as a function of doping. Although the string model and sprinkled-hole simulation both agree with experiment at half-filling by construction, already at low doping the string model performs significantly better than sprinkled holes. The string model is quantitatively accurate across a larger doping range than for $\pi$-flux states, but both are in greater agreement with experiment than the sprinkled-hole simulation. (Bottom) Although the absolute difference between doped and sprinkled-hole pattern-length histograms increases with doping, the shape remains roughly invariant. \textbf{(B)} Average string-pattern length versus doping. Doped AFMs exhibit longer-length string patterns compared to heated AFMs, even when the staggered magnetization or nearest-neighbor spin correlator is equal and holes are sprinkled in to equate doping levels (see text). \textbf{(C)} Total string count at $10\%$ doping as a function of temperature, with corresponding sprinkled-hole string count subtracted. Sensitivity to strings decreases with temperature due to decreased order in the parent AFM as seen in the sprinkled string count (inset). In (A) and (B), error bars on the doping are calculated as in \cite{SI}, section 2.1. All other error bars represent 1 SEM. The figure is based on more than 24,800 experimental realizations.}
	\vspace{-10pt}
\end{figure}

In Fig.~2D and its inset we plot, respectively, the relative and absolute differences between the pattern-length histograms in the doped and undoped cases; these differences are shown for both the cold and hot datasets used in Fig.~2, B and C.
For the undoped case, we use the sprinkled string-pattern length distribution $p_s^\delta(\ell)$ to account for any deviation from the half-filling distribution resulting from the introduction of holes.
Although the absolute difference does not recover the exact analytic string distribution (Fig.~1C), which can be attributed to the imperfect detection of the pattern recognition algorithm, for cold temperatures it does assume a qualitatively similar distribution.
Notably, at $10.0(8)\%$ doping we find more than three times as many length-9 patterns as there are at half filling, reflecting the large impact of holes in an AFM spin background.

Focusing on the cold dataset, we now examine the relationship between doping and the number of detected string patterns (Fig.~3A).
In this string-pattern count, we omit patterns of one or two sites to avoid contributions from quantum fluctuations such as doublon-hole pairs or spin-exchange processes.
The string-pattern count increases with doping and saturates at about $16\%$ doping.
This saturation is consistent with a high density of strings and overlapping or adjacent strings scrambling spin order such that pattern detection becomes insensitive to additional strings.
The continued agreement between geometric strings and experiment in both the string-pattern count and the absolute difference $p^\delta(\ell)-p_s^\delta(\ell)$ suggests that the increase in number of string states is sufficient to explain the experimental data.

The experimental string-pattern count is significantly larger than that of the sprinkled-hole simulation; nonetheless, there is an increase in detected string patterns owing to the additional holes.
The string-pattern count from $\pi$-flux states shows considerably better agreement with experimental data than with sprinkled holes, exhibiting only a slight excess of string patterns at low doping and a deficit at high doping.
The largest deviations occur at low doping, which may be related to the absence of long-range order at zero temperature in $\pi$-flux states at half-filling.

The average string-pattern length quantifies the size of the region around the hole where the spin pattern is distorted by the string (Fig.~3B).
The observed values are comparatively small, influenced by the large contributions from quantum fluctuations at half-filling.
The average string-pattern length does not change substantially with doping, consistent with spatially isolated patterns; however, at larger dopings, we observe a slight decrease in average length that coincides with the observed saturation in the string count.
This behavior is captured by the geometric-string model for low and intermediate doping.
At high doping, the theory exhibits shorter average string lengths than the experiment, which may result from high-string-density effects such as string-string interactions, which are not included in the theory.

We compare these results to a dataset where geometric strings are not expected to occur.
This dataset consists of experimental images taken at various temperatures at half-filling with sprinkled holes to match each desired doping level (for details, see section 3.5 of \cite{SI}).
Temperatures are chosen to match the measured staggered magnetization and capture the observed loss of AFM order.
Notably, the average string-pattern length reveals that this loss through heating occurs in a fundamentally different way than through doping.
For all nonzero doping, the temperature-based dataset exhibits shorter average string-pattern lengths than the experimentally measured doping dataset.
As doping increases, the average length monotonically decreases.
Alternatively, we match the nearest-neighbor spin correlator instead of the staggered magnetization and find an even greater distinction between the doped and temperature-based datasets. 

We better understand the role of temperature in string-pattern detection by observing how the string count varies with temperature at fixed doping.
For $10\%$ doping, we plot the difference between the experiment and sprinkled-hole string counts (Fig.~3C), which are plotted separately in the inset.
At our lowest temperatures, the difference is greatest.
This high sensitivity is consistent with the greatest spin ordering for the parent AFM at low temperatures, accompanied by a relatively large string-pattern count from the experimental data.
The difference decreases steadily with increasing temperature, predominantly owing to the increase in the sprinkled-hole string-pattern count from decreased spin ordering in the parent AFM, vanishing around $T=J$.

\vspace{20pt}
\tocless{\section*{Spin correlations and staggered magnetization}}

An accurate microscopic framework for the Fermi-Hubbard model should also be able to predict more conventional observables such as two-point correlation functions, which have been used with quantum gas microscopes to quantify spin and charge order \cite{Parsons2016, Boll2016, Cheuk2016a, Brown2017}.
To that end, we measure the sign-corrected spin-spin correlation function for displacements $|\mathbf{d}|=d$, averaged over all sites $\mathbf{i}$ in the system and all experimental realizations
\begin{equation}
C_s(|\mathbf{d}|) \equiv (-1)^{||\mathbf{d}||} \frac{\langle \hat{S}_{\mathbf{i}}^z \hat{S}_{\mathbf{i}+\mathbf{d}}^z \rangle - \langle \hat{S}_{\mathbf{i}}^z \rangle \langle \hat{S}_{\mathbf{i}+\mathbf{d}}^z \rangle}{S^2}
\end{equation}
where $\hat{S}^z_{\mathbf{i}}$ is the spin-$S$ operator on site $\mathbf{i}$, $S=1/2$, and $||\mathbf{d}||$ denotes the $L^1$ norm of $\mathbf{d}$, by measuring charge correlations in experimental realizations with and without spin removal \cite{Parsons2016}.
Thanks to the sign correction $(-1)^{||\mathbf{d}||}$, positive correlator values indicate AFM ordering.
Figure 4A shows the nearest neighbor, diagonal next-nearest neighbor, and straight next-nearest neighbor spin correlators ($C_s(1)$, $C_s(\sqrt{2})$, and $C_s(2)$, respectively) as a function of doping at $T = 0.65(4)J$.
At half-filling, $C_s(1)$ is substantially larger than both $C_s(\sqrt{2})$ and $C_s(2)$ due to a strong admixture of spin singlets on adjacent sites \cite{Gorelik2012}.
As the system is doped, all correlators exhibit a reduction in magnitude.
$C_s(1)$ remains positive for all experimentally realized doping values, whereas $C_s(\sqrt{2})$ exhibits a statistically significant sign change around $20\%$ doping.
These features have been observed in experiment \cite{Parsons2016,Cheuk2016a, Koepsell2018} and numerics \cite{Cheuk2016a}, and are good benchmarks for the evaluation of theoretical models.

\begin{figure}
	\begin{center}
		\includegraphics[width=\columnwidth]{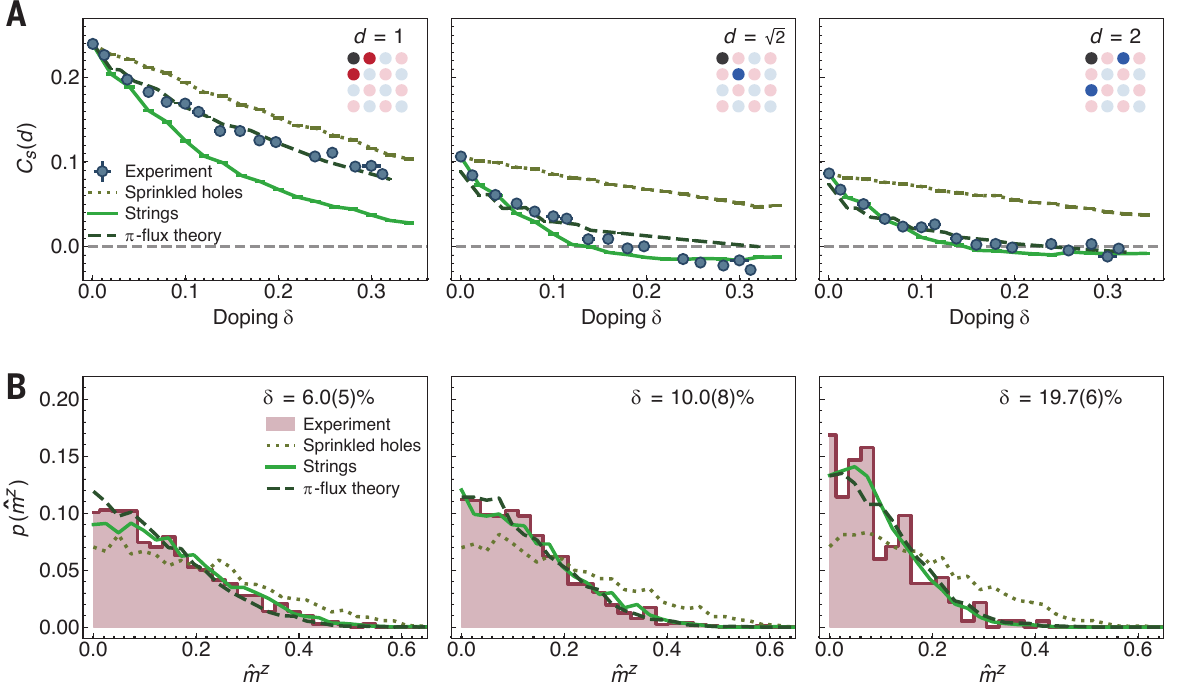}
		\label{fig4:spin}
	\end{center}
	\vspace{-25pt}
	\caption{\textbf{Spin correlations and staggered magnetization. (A)} Decay of nearest-neighbor (left), diagonal next-nearest-neighbor (center), and straight nearest-neighbor (right) spin-spin correlation functions upon doping. The $\pi$-flux theory most quantitatively explains $C_s(1)$, but only the string model captures the sign change of $C_s(\sqrt{2})$. In all three cases, sprinkled holes overestimate the spin correlations. Doping error bars are calculated as in \cite{SI}, section 2.1; all other error bars represent 1 SEM. \textbf{(B)} Full counting statistics of the staggered magnetization for doping values of $6.0(5)\%$ (left), $10.0(8)\%$ (center) and $19.7(6)\%$ (right). Both $\pi$-flux states and geometric strings show reasonable agreement, whereas sprinkled holes do not. The figure is based on more than 29,900 experimental realizations at average temperature $T=0.65(4)J$.}
	\vspace{-10pt}
\end{figure}

We make predictions for spin correlations from ensembles of non-postselected images with sprinkled holes, geometric-string theory, or $\pi$-flux states.
By construction, at half-filling the predictions of sprinkled holes and the string model are the same as those of experimental half-filling data.
Away from half-filling, sprinkled holes underestimate the decrease of the correlators because the model fails to account for the disruption of AFM order as the system is doped.
By contrast, beginning at intermediate doping values, the string model overestimates the decrease of $C_s(1)$, which could stem from backaction of the background state after string-state formation.
However, it explains the decrease of $C_s(\sqrt{2})$ and $C_s(2)$ on a quantitative level.
The $\pi$-flux model performs well and accurately predicts $C_s(1)$ and $C_s(2)$ far from half-filling but fails to predict the sign change of $C_s(\sqrt{2})$ at intermediate doping, even when the fitted temperature is varied.
The sign change of $C_s(\sqrt{2})$ is an interesting qualitative feature that is predicted and can be explained by the string model.
As a direct result of spins being displaced by one site when a string passes through, $C_s(1)$ is mixed into $C_s(\sqrt{2})$.
Because $C_s(1)$ reflects opposite spin alignment from $C_s(\sqrt{2})$, this mixing results in a sign change once the contribution of $C_s(1)$ exceeds that of the original correlation strength at some critical doping.

Cold-atom experiments provide access to full-counting statistics (FCS) because of their ability to project and measure an entire quantum system at once \cite{Mazurenko2017}.
We measure the FCS of the staggered magnetization operator
\begin{equation}
\hat{m}^z = \frac{1}{N} \sum_{\mathbf{i}} (-1)^{||\mathbf{i}||} \frac{\hat{S}_{\mathbf{i}}^z}{S}
\end{equation}
for system size $N$ across all experimental realizations as we dope the system (Fig.~4B).
As expected, the staggered magnetization distribution narrows, reflecting the finite-size crossover from the AFM-ordered phase \cite{Mazurenko2017}.
The sprinkled-hole simulation does not exhibit a major change in the distribution as the system is doped, as it fails to account for holes disrupting the AFM order.
By contrast, both $\pi$-flux states and geometric strings demonstrate reasonable agreement with the experimentally measured distribution function across all dopings.
Across all observables considered, both of these theories perform quite well, especially in comparison to the sprinkled-holes simulation and the na\"ive phenomenological models detailed in section 5 of \cite{SI}.
However, we find the sign change of $C_s(\sqrt{2})$ to be a key qualitative feature that is captured only by geometric strings.

\vspace{20pt}
\tocless{\section*{Antimoment correlations}}

All observables studied in this work thus far have focused on the spin sector of the Hubbard model.
Next, we examine correlations in the charge sector.
At sufficiently low temperatures, one may expect signatures of pairing \cite{Emery1995, Keimer2015} or stripe phases \cite{White1998, Zaanen2001}, which lead to hole bunching.
However, anticorrelations of the holes, as observed previously at increased temperatures \cite{Cheuk2016a}, are expected in the strongly correlated metallic regime of the Hubbard model.
The transition between these two regimes in the Hubbard model phase diagram is not yet fully understood; however, the currently accessible experimental regime allows us to place new bounds on where this transition can occur.
We continue to compare to predictions of $\pi$-flux states, but do not compare to predictions of the geometric string theory because it approximates that charges are uncorrelated.
Rather, because each string is associated with a single hole, correlation functions of holes can reveal possible interactions and correlations between geometric strings, should they exist.

In our experiment, doubly-occupied sites appear as empty when imaged and the exact hole correlation is not directly accessible; rather, we measure ``antimoment" correlations $C_h(|\mathbf{d}|)$ at a distance $|\mathbf{d}|$, which include contributions from doublon-doublon and doublon-hole correlations:
\begin{multline}
C_h(|\mathbf{d}|) \equiv
	\Big(
		\left\langle
			\left(1-\hat{n}_{s, \mathbf{i}}\right)
			\left(1-\hat{n}_{s, \mathbf{i+d}}\right)
		\right\rangle - \\
		\left\langle
			\left(1-\hat{n}_{s,\mathbf{i}}\right)
		\right\rangle
		\left\langle
			\left(1-\hat{n}_{s,\mathbf{i+d}}\right)
		\right\rangle
	\Big)
\end{multline}
where $\hat{n}_{s,\mathbf{i}}$ is the single particle occupation on site $\mathbf{i}$.
Note that this correlator is identical to the moment correlator.
At half-filling, numerics indicate positive antimoment correlations at the percent level for nearest neighbors, dominated by positive doublon-hole correlations \cite{Cheuk2016a}.
Doublon-hole pairs beyond nearest-neighbors become increasingly unlikely; therefore, to avoid the effects of doublon-hole pairs, we focus on correlations at distances greater than 1.
We find the nearest-neighbor antimoment correlator at half-filling to be weaker than predicted according to numerics, which may result from imperfect imaging fidelity.
However, this effect only weakens the magnitude of the antimoment correlators measured; we therefore focus on qualitative conclusions from the experimental data.

\begin{figure}
	\begin{center}
		\includegraphics[width=\columnwidth]{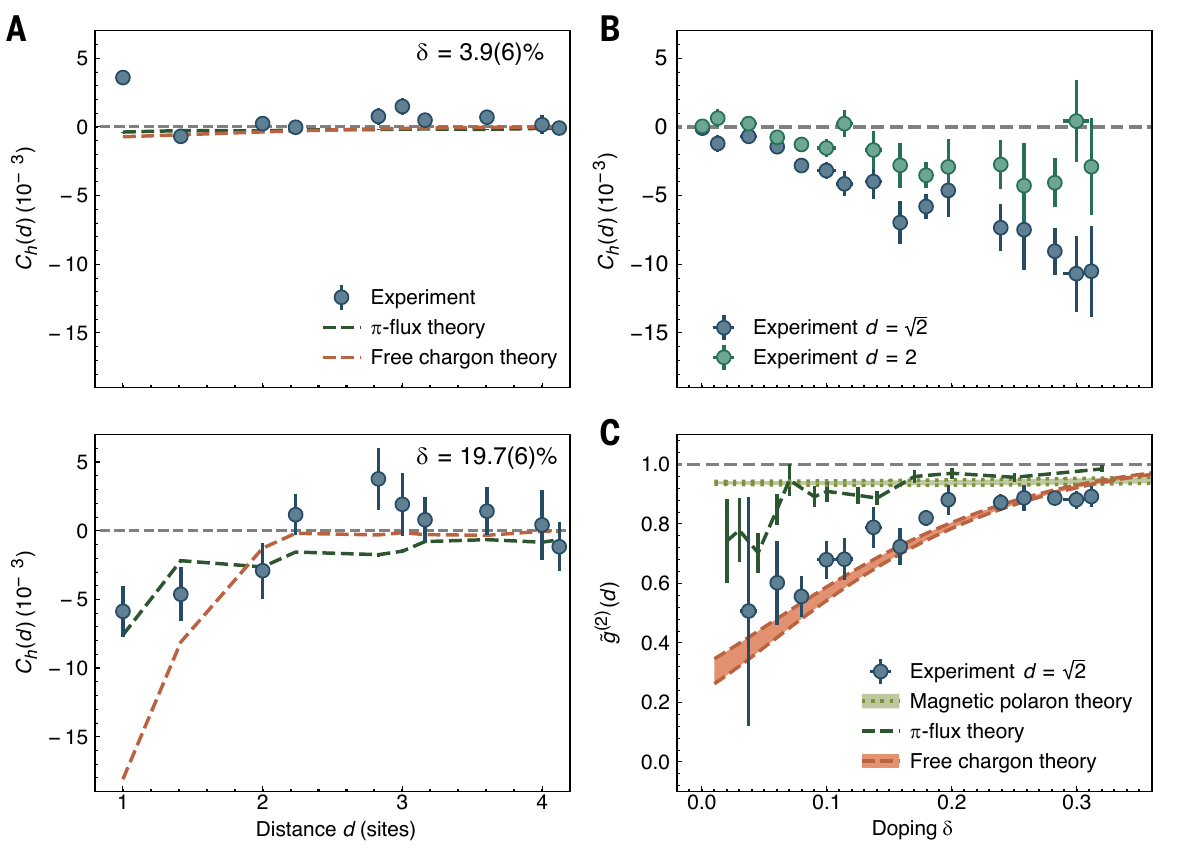}
		\label{fig5:spin}
	\end{center}
	\vspace{-25pt}
	\caption{\textbf{Observation of hole antibunching. (A)} Antimoment correlation function for weak (top) and strong (bottom) doping. The correlation functions are different up to a distance of $d=2$. \textbf{(B)} Diagonal next-nearest neighbor and straight next-nearest neighbor antimoment correlators versus doping. At both distances negative correlations grow with doping. \textbf{(C)} Normalized antimoment correlator at $d = \sqrt{2}$ versus doping. The experimental result cannot be explained by the $\pi$-flux or a point-like magnetic polaron theory (see text), but instead matches a free fermionic chargon theory. In (B) and (C), error bars on the doping are calculated as in \cite{SI}, section 2.1. All other error bars represent 1 SEM. The figure is based on more than 9900 experimental realizations at an average temperature $T=0.65(4)J$.}
	\vspace{-10pt}
\end{figure}

Figure 5A shows the antimoment correlation for $3\%$ (top) and $19\%$ doping (bottom) at a temperature $T=0.65(4)J$.
Whereas holes appear uncorrelated close to half-filling, at larger doping qualitatively different behavior appears.
We find statistically significant antimoment anticorrelations out to distances over two sites, reflecting hole-hole repulsion in this regime.
Microscopically, such repulsive interactions can arise from the existence of a low-lying bound state of two holes \cite{Chin2010}.
Here we do not consider geometric-string theory or sprinkled holes because both introduce uncorrelated holes by construction.
Additionally, in the comparison to $\pi$-flux states, we do not include doublon-hole pairs to avoid unintended artifacts in the antimoment correlator.
For reference, we plot the predicted hole-hole correlation function for a phenomenological model of spinless fermionic chargons with nearest-neighbor hopping of strength $t$ and temperatures between $0.5J$ and $0.7J$ \cite{Kaul2007}.
Here, strong anticorrelations result from Pauli repulsion between the fermionic chargons, but qualitatively similar behavior is expected for bosonic chargons with hard-core interactions.
We find that both theories qualitatively describe the experimental result.

The emergence of this repelling behavior can be characterized by plotting the antimoment correlation as a function of doping for $d = \sqrt{2}$ and $d = 2$ (Fig.~5B). Beyond the intermediate doping regime, negative correlations appear at distances of $\sqrt{2}$ and $2$, suggesting a growth of hole-hole repulsion with doping. Furthermore, the presence of antimoment correlations between sites of differing sublattices at $d = 1$ evidences against holes tunneling preferentially between sites of one sublattice, as predicted by theories of pointlike magnetic polarons with a dispersion minimum at $(\pi/2, \pi/2)$ in the Brillouin zone \cite{Kane1989, Sachdev1989, Martinez1991, Liu1992}.

Finally, we plot a normalized $g^{(2)}(d=\sqrt{2})$ to account for the difference between doped holes and holes in doublon-hole pairs and quantify the relative fraction of doped holes that are anticorrelated:
\begin{equation}
\tilde{g}^{(2)}(|\mathbf{d}|) \equiv \frac{C_h(\mathbf{d})}{\delta^2} + 1
\end{equation}
for doping $\delta$ (Fig.~5C).
This rescaling allows direct comparison to the $g^{(2)}$ function for theories without doublon-hole pairs.
The number of free holes is too small for doping below $5\%$ to make statistically significant statements about the behavior of holes in this regime.
In the geometric-string theory, we assume that chargons (dressed dopants) are completely uncorrelated with each other, but because of their fermionic statistics, Pauli blocking should actually introduce anti-correlations which have not yet been included in our analyses.
We first consider a description of these chargons as pointlike magnetic polarons, where the known dispersion relation of the dressed hole \cite{Brunner2000} is used to define a tight-binding hopping model of the polaron.
Figure 5C shows that our data are incompatible with this model, which predicts significantly weaker hole-hole anticorrelations.
Similar behavior is predicted by the $\pi$-flux theory, which models the doped holes as pointlike objects moving in a quantum spin liquid of singlets. 

Next, we examine a picture of free chargons, motivated by considering magnetic polarons with a finite extent that results from the spinon-chargon bound state predicted by geometric-string theory.
At sufficiently large chargon density, or doping, the chargons are expected to interact, and their hard-core character will introduce anticorrelations.
In this regime, geometric strings are also expected to overlap substantially and modify the dispersion relation of the chargons to be independent of spinons, yielding spinless chargons.
We find that the experimental results demonstrate consistency with these free chargons, in agreement with earlier theoretical work in the strange-metal regime \cite{Kaul2007, Sachdev2016}.
This extension to the geometric string theory may be able to explain the deviations from experiment seen in other observables at high doping, but additional analyses are required.

\vspace{20pt}
\tocless{\section*{Conclusions and outlook}}

The string-pattern-based observables introduced here complement established observables such as correlation functions or full counting statistics.
Across the observables considered, we find better agreement with experimental data between both the geometric-string theory and $\pi$-flux states, as compared to sprinkled holes.

At intermediate doping values, we find evidence for hole-hole repulsion.
Although signatures of other phases such as stripe phases, incommensurate spin order, or nematic fluctuations have not yet been observed in this system, they are predicted to emerge at lower temperatures.

The ideas presented can be extended to other real-space patterns, for example patterns that reflect the underlying physics of other candidate microscopic theories for the doped Hubbard model.
Moreover, machine learning techniques could be used to directly compare sets of raw experimental atom distributions to theoretical models without the need for intermediate observables \cite{Bohrdt2019}.
This class of techniques is highly promising as quantum simulations of the Hubbard model continue to probe lower temperatures within the pseudogap and strange-metal phases, but can also be applied to spatially resolved studies of quenches across phase transitions \cite{Bernien2017}, dynamical phase transitions \cite{Zhang2017}, and higher-order scattering processes \cite{Feng2018}.
Possible extensions of our work include systems with anisotropic spin interactions \cite{Grusdt2018} or doped SU($N$) spin models \cite{Honerkamp2004}.

\tocless{}

\tocless{\section*{Acknowledgments}}
We thank M.~Kan\'asz-Nagy for Heisenberg QMC code.
We thank A. H\'ebert, S. Sachdev, Z.-Y. Weng, and J. Zaanen for insightful discussions.
We acknowledge support from AFOSR; DoD NDSEG; the Gordon and Betty Moore Foundation; NSF; ONR; SNSF; Studienstiftung des deutschen Volkes; and the Technical University of Munich - Institute for Advanced Study, funded by the German Excellence Initiative and the European Union FP7 under grant agreement 291763, the DFG under Germany's Excellence Strategy--EXC-2111--390814868, through DFG grant no. KN 1254/1-1, and DFG TRR80 (Project F8).
All experimentally measured site-resolved atom distributions and analysis code are available \cite{Chiu2019}.

\clearpage
\onecolumngrid

\setcounter{section}{0}
\setcounter{subsection}{0}
\setcounter{figure}{0}
\setcounter{equation}{0}

% Usual (decimal) numbering
\renewcommand{\thesection}{\arabic{section}}
\renewcommand{\thesubsection}{\thesection.\arabic{subsection}}
\renewcommand{\thesubsubsection}{\thesubsection.\arabic{subsubsection}}

% Fix references
\makeatletter
\renewcommand{\p@subsection}{}
\renewcommand{\p@subsubsection}{}
\makeatother
\renewcommand{\numberline}[1]{#1~}

\makeatletter
\renewcommand{\bibnumfmt}[1]{[S#1]}
\renewcommand{\citenumfont}[1]{S#1}
\renewcommand{\thefigure}{S\@arabic\c@figure}
\makeatother
\renewcommand{\theequation}{S\arabic{equation}}

	\begin{center} \begin{Large}
		Supplementary Materials for\\ \textbf{String patterns in the doped Hubbard model}
	\end{Large} \end{center}
	
	\baselineskip 16pt
	\tableofcontents
	
	\newpage
	
	\baselineskip24pt
	
	\section{Atomic sample preparation}
	Details about the experimental setup, including the procedures used to create the low-temperature Fermi gas and set the doping value, can be found in \cite{Mazurenko2017}. The temperature of the gas is increased via the process described in \cite{Greif2016}. For all measurements presented, $U/t = 8.1(2)$ and is calibrated as described in \cite{Chiu2018}.
	
	\section{Doping determination} \label{doping_determination}
	In our experiment, we measure the percentage of sites occupied by single particles (singles density). We use numerical simulations to obtain the doping as a function of the singles density. For data between $T=0.6J$ and $T=0.8J$ we use data obtained from a determinantal quantum Monte Carlo algorithm \cite{Varney2009, Brown2019}, and for all larger temperatures we use data obtained from a numerical linked-cluster expansion algorithm \cite{Khatami2011}. For $T < 0.6J$, the sign problem becomes significant. As a result, in this regime we use data at $T = 0.6J$, as the density sector of the equation of state is relatively insensitive to temperature here. We account for an imaging fidelity of $98.5\%$. When statistical fluctuations cause the singles density to exceed the numerically-obtained singles density at half-filling, we treat those samples as at half-filling.
	
	\subsection{Error analysis}
	When determining the standard error of doping values for each experimental dataset, we assume that the particle density is linearly dependent on singles density. We apply a linear fit to doping versus singles density from the numerical simulation mentioned above, yielding approximately $\delta = 1.22 \times (0.905 - n_{s})$, where $\delta$ is doping and $n_{s}$ is the singles density. We then calculate the standard error of the singles density and use the linear fit result to get the standard error of the mean doping value.
	
	Since the actual doping value varies across datasets, we group datasets by their mean doping values within windows of width $2\%$. This yields a single mean doping value $\bar{d}$ for the entire group. The associated uncertainty $\Delta$ is determined by assuming each dataset $k$ within the group was taken at a different doping value $d_{k}$ with a corresponding uncertainty $\delta d_k$. Then $\Delta$ can be calculated as:
	\begin{equation}
	\Delta = \sqrt{\frac{1}{\sum_{k} n_{k}}\sum_{k} ((d_{k}-\bar{d})^{2}+{\delta d_{k}}^{2})n_{k}}
	\end{equation}
	
	For datasets which are sufficiently close to half-filling, fluctuations of additional holes or particles will both result in a decrease of the singles density. This single-sided cut-off of statistical fluctuations will lead to a systematic offset in the mean. To estimate this offset, we assume that the statistical fluctuations in total density follow a normal distribution centered at half filling with standard deviation $\sigma$. Then the resulting distribution in the singles density follows a half-normal distribution, characterized by an offset in the mean of $\sigma\sqrt{2/\pi}$ and standard deviation of $\sigma \sqrt{1-2/\pi}$. Note that this provides an upper bound of the systematic offset, because in reality our datasets are not all centered exactly at half filling. If we consider all datasets which are within one standard deviation of half filling, this results in an estimated systematic offset in the mean doping of the grouped dataset of about $0.25\%$. This systematic uncertainty is included in the errorbar for half-filling doping values. We note that a higher-order correction to the dependence of the singles density on the total density makes the singles density less sensitive and therefore would only decrease the magnitude of this systematic error.

	\section{String pattern detection}
	
	\subsection{Algorithm}
	
	See Fig.~S1 for a detailed schematic for the string pattern detection algorithm. It consists of three main steps: (i) postselection, as described in the main text; (ii) determination of sites which deviate from a checkerboard pattern, and (iii) extraction of string patterns from those sites which deviate, according to the rules described in the main text. Here we elaborate on the implementation of these steps.
	
	In (i), we calculate the staggered magnetization of a circular region (``window") of diameter 7 sites (as shown in Fig.~S1) as the region is scanned over the entire 10-site-diameter sample. For each image, we use the 7-site-diameter window of highest staggered magnetization. If there are multiple such possible regions, we take the upper-left-most one, however this is an arbitrary choice given that the entire sample is homogenous. Once all images have been reduced in size, we postselect on all data for the top $60\%$ of the staggered magnetization. We discuss the robustness to postselection and finite-size effects in section 3.3.1 and Fig.~S2.
	
	In (ii), of the two possible checkerboard patterns, we select the one closer in sublattice magnetization as reference for each image separately.
	
	In (iii), we first sort the sites which deviate into disjoint sets, each of which consists of sites which can be connected by nearest neighbors. For each set, we identify all empty sites, as these may be the end of a geometric string. For each of these empty sites, we trace out all possible strings (sites which are connected via nearest neighbors to at most two other sites) and select the longest one. If there are multiple longest strings, we select one with the upper-left-most starting site. Again this is an arbitrary choice given the homogeneity of the system; we find that modifying this bias does not affect the resulting string-pattern length distribution. The sites which are part of the identified string pattern are then removed from the set and the process is repeated until no more patterns can be found, i.e. the set no longer contains empty sites. This process is then repeated for each of the disjoint sets.
	
	We note that overlaps of strings or loops within strings are not treated correctly, because sites that do not deviate from the reference state are not taken into account, however given the readout in the Fock basis other algorithms will be similar in this regard. We discuss different detection algorithms in section 3.4 and Fig.~S3.

	\subsection{Total string count}
	
	For low doping, we can fit the data of Fig.~3A to a line to estimate a string detection efficiency for strings of lengths greater than two sites. We find a slope of $1.7(2)\times10^{-3}$ string patterns per site per percent doping for doping up to $6\%$.  The analytically calculated string length distribution for a temperature of $0.6J$ predicts that $65\%$ of string states have length greater than 2 sites, giving an approximate detection efficiency of $25(2)\%$. 
	
	In Fig.~S7A we include all string pattern lengths in computing the total string count, rather than omitting string patterns of length one or two sites as in the main text. We find that all simulations show similar agreement with experimental data, and all string counts increase linearly with doping, reflecting that this quantity may simply reflect the doping level.

	\subsection{Algorithm evaluation}
	
	\subsubsection{Postselection and finite-size effects}
	
	The size of the postselection region is chosen to be 7 sites in diameter according to the AFM correlation length at half filling. We vary the window to a smaller circular region of 5 sites in diameter, or to a larger circular region of 8 sites in diameter, and find that the qualitative dependence of the string count on doping remains the same, see Fig.~S2A. The baseline string count at half filling increases for larger postselection regions as the region becomes larger than the correlation length and the deviation from the reference checkerboard increases. 
	
	This finite-size variation seems to affect experiment and all theoretical simulation results similarly, which is reasonable given that the pattern detection algorithm is identical in all analyses. However, in Fig. S7G, we simulate strings in the infinite-length limit and detect a greater number of string patterns upon doping. This result suggests that when the system is much smaller than the length of the string patterns, the number of detected strings is biased higher. Indeed, increasing the size of the postselection region results in a slight decrease of the estimated detection efficiency.
	
	We also consider the effect of moving the window to achieve the highest values of the staggered magnetization. We find that fixing the window to the center of the system while keeping a postselection threshold of $60\%$ greatly increases the number of string patterns found at half filling relative to the additional number of patterns found upon doping the system, see Fig.~S2B. This is due to a greater average deviation from the reference checkerboard. In principle, we could achieve a better signal to noise by postselecting more strongly on which images we use, at the cost of increased statistical fluctuation. While the half-filling value changes, the estimated detection efficiency does not change statistically significantly, indicating robustness of the detection algorithm to this effect.
	
	Finally, the fraction of images kept in the post-selection process can be varied. We choose to keep the top $60\%$ of images in an effort to capture the tail of the histogram of the staggered magnetization, while maintaining a reasonably high number of images. Upon changing the postselection to $40\%$ or $80\%$, see Fig.~S2C, we find fewer or more string patterns at half-filling, respectively. However, the slope of the string count as a function of doping in the low-doping regime does not change statistically significantly.

	\subsubsection{Comparison to full readout}
	
	In images taken in our experiment, we do not distinguish between holes, doublons, and the removed spin species. In a system with full readout, this distinction is available. In this case, the hole positions are known and the number of detected string patterns must correspond to the number of holes which are not in doublon-hole pairs, i.e. dopants. However, the detected distribution of string lengths can still be modified by overlaps between strings in the same way as in our experiment. We simulate full spin readout in quantum Monte Carlo simulations of the Heisenberg model with simulated strings. The simulation is performed on a 40-site by 40-site system with periodic boundary conditions, from which a 10-site-diameter disk is cut out to match the experimental system. Postselection is then done in the same way as in the experimental data analysis. In Fig.~S4 a comparison of the detected string length distribution with and without full readout is shown, where the distribution obtained without full readout has the half-filling distribution subtracted. While the signal with full readout is a factor of about five higher, the relative distribution of the detected string-pattern lengths remains the same.
	
	\subsubsection{Signal at half-filling}
	
	Apart from doublon-hole pairs, the Fermi-Hubbard model at half filling for $U \gg t$ can be approximated by the Heisenberg model. We can therefore examine the detected string-pattern length distribution from Heisenberg QMC simulation to better understand our experimental signal at half filling. For consistency, after simulating a 40-site by 40-site system with periodic boundary conditions, we cut out the same sample size and use the same readout and post-selection schemes as in the experiment, and we add doublon-hole pairs into the simulation by converting neighboring sites with opposite spins into doublon-hole pairs with a probability given by $4t^2/U^2$. 
	
	Fig.~S5 shows the string length distribution from the experiment at half filling as well as from QMC simulations of the Heisenberg model at $T = 0.6J$ with and without artificial doublon-hole pairs. The introduction of artificial doublon-hole pairs corrects the significant discrepancy between the QMC data and the experimental data at strings of short lengths. The resulting simulated data agrees reasonably well with experiment, suggesting that the detected string patterns in the experimental data at half filling come from the deviation of a quantum AFM from the checkerboard reference pattern. Slightly more long string patterns are found in Heisenberg QMC snapshots compared to experiment half-filling; this discrepancy may be due to a failure to consider non-adjacent doublon-hole pairs, which are expected at the experimental parameters. Indeed, this effect decreases upon doping as doublon-hole fluctuations become rarer, where the QMC data with simulated strings agrees quite well with the experimental measurements.
	
	\subsubsection{Temperature dependence}
	
	In Fig.~3C of the main text we show that for a doping of $10\%$, the pattern detection algorithm is only sensitive below a temperature $T=J$. We extend this analysis and plot the total string count as a function of temperature for different doping values in Fig.~S6B. The temperature at which saturation occurs decreases with increased doping, demonstrating that string count saturation occurs through the combination of doping and temperature-dependent background spin order. This effect may be exacerbated by the increase in string length with temperature. According to the geometric string theory, while the string length distribution predicted by analytic calculations is dominated by strings of length 0 and 1 for temperatures smaller than $0.5J$, it continually broadens and longer strings are more likely to appear for increasing temperature, see Fig.~S6A.
	
	\subsection{Alternate algorithms}
	
	There are many possible algorithms which can be used to quantify the presence of string patterns. Here we discuss two alternate algorithms. We find that these algorithms are comparable in performance to the detection algorithm discussed in the main text, and determine our algorithm of choice based on simplicity while making use of all information available.
	
	\subsubsection{Simplified difference method}
	The most straightforward way to detect string patterns is to simply count the continuously connected sites that deviate from the classical checkerboard pattern. As opposed to the algorithm we use in the main text, not every object identified as a string pattern in this way can actually be a geometric string. For example, it is possible that both endpoints as well as the sites surrounding them are occupied such that there cannot be a hole at either end. Moreover, the shape of the object may not be consistent with a non-branching string pattern. However, one can argue that these inaccuracies mainly occur at high temperature or high doping values when perturbations and strings start to overlap. 
	
	In Fig.~S3A, the same quantities as plotted in main text Figs.~2B and 3A are shown, but under the simplified string detection algorithm instead. At a doping of about $10\%$, both $\pi$-flux states and geometric strings seem to quantitatively match the experimentally measured string-pattern length distribution well. The total string count versus doping looks qualitatively similar compared to that of the algorithm used in the main text. The detection efficiency remains roughly the same as before, while the half-filling baseline is slightly larger. 
	
	\subsubsection{Happiness method}
	In the dilute string regime, where string states do not overlap or lie adjacent to one another, one can search for string patterns by also requiring that sites immediately surrounding the string maintain AFM order. Note that this requirement also omits string states which have segments that lie adjacent to each other, for example sting patterns containing a tight ``U"-shape. This method is also susceptible to identifying string patterns caused by doublon-hole pairs, spin-exchange processes, and projective measurement. However, as these effects will introduce deviations from AFM order, this is perhaps the most conservative approach to finding string patterns. 
	
	This algorithm characterizes nearby order by labeling each site with the number of anti-aligned bonds it has with its nearest neighbors, termed the "happiness" of that site, for images with one spin species removed. For example, sites in a classical AFM would all be labelled with happiness 4, while a ferromagnet would have sites with happiness 0. As a hole moves through an AFM, sites which previously had happiness 4 will exhibit reduced happiness. Depending on the length of the string, sites within a string will have specific happiness values. Based on this, the algorithm takes images with one spin state removed and for each image, begins by storing all sites which could be the beginning of a string. For each candidate string beginning, it sees if there is a neighboring site that could be the next site in the string, given the happiness and spin occupation of that site. This process continues until the string cannot be propagated any further, at which point the algorithm searches for a neighboring site which could be the end of the string.
	
	Figure S5B shows the same quantities plotted in main text Figs.~2B and 3A, but under the happiness string detection algorithm instead. Note that the signal to noise ratio is significantly lower and the absolute signal itself is lower by almost an order of magnitude. This is not surprising, especially given that quantum fluctuations and projection noise do contribute considerably to measurement and reduce the sensitivity of string patterns to string states. Here the experimental result seems to best match the sprinkled holes simulations, however uncertainties are large and this is highly inconsistent with all other results with conventional observables, requiring further investigation.

	\subsection{Average string length}
	
	The average measured string length $\overline{l}(\delta)$ in Fig.~3B of the main text is calculated from the string histograms
	\begin{equation}
	\overline{l} = \sum_l l \cdot p^{\delta} (l) / \sum_l p^{\delta} (l). 
	\end{equation}
	Error bars are obtained via standard error propagation of the uncertainties in the measured string length probabilities. 
	
	As discussed in the main text, we obtain the average string length for datasets where geometric strings are not expected to occur by using experimental datasets taken at various temperatures and half filling (``temperature datasets"). Then, for each desired doping value we randomly place holes into all temperature datasets to artificially achieve the doping value for every temperature dataset. We then extract the staggered magnetization and average string length of each dataset, obtaining the relationship between these two quantities. We perform a linear fit to obtain $\overline{l}_\delta(m_z)$, the average string length at a given doping as a function of staggered magnetization, for the temperature datasets.
	
	To determine which value of $m_z$ to use in this function, we use the low-temperature experimental datasets taken at low temperature and various dopings (``doping datasets"). As the dependence of the measured staggered magnetization on doping is non-linear for these datasets, we perform a linear fit of the closest five data points for each doping value to obtain a reliable estimate of $m_z$. This value is then used to determine the average string length for the temperature datasets. The entire process is repeated for each doping value to obtain the data in Fig.~3B of the main text (orange circles).
	
	This procedure allows us to directly compare the experimental data at finite doping to a scenario where the staggered magnetizations are similar and the same number of doped holes are present - but no geometric strings are included. The error bar for the predicted average string length is obtained by combining the measurement error of $m_z$ with the error of the linear fits weighted by the standard deviation of the measured quantities. As a cross-verification, we have applied the same procedure except for choosing the 'effective' temperature by matching the value of the nearest-neighbor spin correlator. With this method we found the same qualitative behavior, see Fig.~3B of the main text (red circles).

	\section{Geometric string simulation}
	
	Given the background signal and imperfect detection efficiency of the pattern finding algorithm, we cannot directly compare the measured string length distribution to the predictions from geometric-string theory. We therefore instead use these predictions to simulate snapshots, which in turn are analyzed to obtain string length distributions which can be directly compared. Because this theory makes no statement about the parent AFM, in this simulation a number of holes corresponding to the desired doping value are placed at random positions into the experimental images taken at half filling. For each hole, a length is sampled from the analytic string length distribution and the hole is propagated accordingly. The direction is chosen randomly at each step, but the hole cannot move backwards. This procedure produces a set of images which are then analyzed identically to the experimental dataset, such that the pattern detection scheme is common to both.

	\subsection{Robustness to changes in geometric string predictions}
	
	We vary the geometric string theory prediction of the string length distribution to examine how the resulting detected string-pattern length distribution changes, see Fig.~S7B-G. Changing the temperature for Boltzmann sampling of the string states yields a worse agreement with the experimental result, as does changing the participation ratio of holes in strings by only moving a fraction of the holes which have been randomly placed. We also alternatively select only strings of a given length and find the best agreement for strings of length 4; this is close to the average string length at $T=0.6J$ of 4.2. From these results we conclude that perturbations to the analytic string length histogram are unlikely to improve agreement with the experimental measurement. We note that the decrease in string count at very high doping for infinite-length strings is likely an artifact from simulation.

	\section{Phenomenological models}
	
	While the results of the main text indicate that both geometric strings and $\pi$-flux states predict experimental result better than sprinkled holes, it is also constructive to assess how well basic phenomenological models perform. Here we consider two.
	
	\subsection{Matching spin correlations}
	Here we begin with a random but balanced spin distribution. From this ensemble, we randomly place the desired number of doublon-hole pairs and holes according to the desired doping value. Finally, we flip spins randomly until the correlators $C_s(1)$ and $C_s(\sqrt{2})$ agree with the experimental data. From this dataset, we apply our string pattern detection algorithm to compare with experimental result. The region of interest of the dataset matches that of the experiment. We generate images corresponding to half-filling and to $10\%$ doping in experiment.
	
	Fig.~S8A-C shows the measured string pattern length distribution, spin correlation function, and full counting statistics of the staggered magnetization for the generated images in comparison to experimental result. Because we begin with spin distributions with no correlations and artificially introduce nearest-neighbor and diagonal next-nearest neighbor correlations, it is not surprising that the correlation functions do not agree beyond short distances. In turn, because the spin correlation function at large distance is closely related to the average staggered magnetization, it is not surprising that the staggered magnetization distribution also does not agree and that the average value is lower for the generated data.
	
	However, the string pattern length distributions do not match either. While there is agreement at short string pattern lengths, the generated images contain statistically significantly more long patterns than in the experiment, especially for half-filling. Surprisingly, it seems that matching the first two correlators is insufficient to introduce the order needed to prevent long string patterns. Modifications to the phenomenological model such as beginning from a perfect checkerboard pattern with SU(2) symmetry do not increase the level of agreement. Furthermore, the additional contributions of making $C_s(\sqrt{2})$ match experiment (as compared to just $C_s(1)$) are small.
	
	\subsection{Corrections to a classical AFM}
	We also apply a phenomenological approach where we begin with a classical checkerboard, create singlets with some variable density, and place doublon-hole pairs and holes randomly according to the desired doping value. We finally apply a projective measurement process and, ensuring that the region of interest is the same as in experiment, run the string search algorithm on the result.
	
	Fig.~S8D-F show that while the density of singlets can be varied to achieve reasonable agreement for the staggered magnetization full counting statistics and string-pattern length distribution at $10\%$ doping, the corresponding spin correlation function seems unphysically flat at distances beyond the nearest neighbor. Furthermore, it is clear that keeping the same density of singlets for half-filling results in stark disagreement across all observables.

	\section{Theoretical background}

	\subsection{Geometric-string theory}
	
	To describe the effect that hole doping has on the AFM at half filling, we neglect correlations between dopants and consider the case of a single hole. Our starting point is the undoped Heisenberg spin model at half filling, which we describe by a thermal density matrix $\rho_{1/2} = e^{- \beta \hat{\mathcal{H}}_{\rm H}} / Z_{1/2} $, where $\beta = 1 / k_{\rm B} T$ with the Boltzmann constant $k_{\rm B}$ and temperature $T$, $Z_{1/2}$ is for normalization and $ \hat{\mathcal{H}}_{\rm H} = -J \sum_{\langle \mathbf{i}, \mathbf{j} \rangle} \hat{\mathbf{S}}_{\mathbf{i}} \cdot \hat{\mathbf{S}}_{\mathbf{j}} $ denotes the Heisenberg Hamiltonian with coupling $J$ between spins $\hat{\mathbf{S}}$ on neighboring sites $\mathbf{i}$, $\mathbf{j}$ of a square lattice. When modeling the correlations between the mobile hole and the surrounding spins, we apply the frozen-spin approximation introduced at zero temperature in Refs. \cite{Grusdt2018,Grusdt2018a}. To describe the motion of the hole, we introduce an approximate basis generated by string states. For example, the trivial string state $| \mathbf{j}, \sigma, 0 \rangle$ with length $\ell=0$ and spin $\sigma$ is obtained by annihilating a fermion with spin $\sigma$ at some lattice site $\mathbf{j}$, i.e. $ | \mathbf{j}, \sigma, 0 \rangle = \hat{c}_{\mathbf{j},\sigma} | \Psi_{1/2} \rangle $, where $| \Psi_{1/2} \rangle $ denotes any typical undoped state from the ensemble described by $\rho_{1/2}$. Non-trivial strings $\Sigma$, defined as finite trajectories on a square lattice without self-retracing components, correspond to sites on a fractal Bethe lattice, or a Cayley tree, with coordination number $z=4$. Every such string labels a separate approximate basis state, $| \mathbf{j}, \sigma, \Sigma \rangle = \hat{G}_\Sigma | \mathbf{j}, \sigma, 0 \rangle$; the string operator $\hat{G}_\Sigma $ starts from the original position $\mathbf{j}$ of the hole and moves it along the trajectory described by $\Sigma$, while displacing all spins along the way accordingly. 
	
	If $| \Psi_{1/2} \rangle $ is the classical N\'eel state, the string states $| \mathbf{j}, \sigma, \Sigma \rangle $ form an orthonormal basis, except for certain loop configurations which have been identified first by Trugman \cite{Trugman1988} and lead to double counting of some states. As shown in Ref.~\cite{Grusdt2018} however, one may assume that all states $| \mathbf{j}, \sigma, \Sigma \rangle $ are mutually orthonormal; the dominant effect of Trugman loops can be captured by adding corrections to the hole dispersion. If $| \Psi_{1/2} \rangle $ describes the ground state of the quantum Heisenberg AFM, the approximation that all states $| \mathbf{j}, \sigma, \Sigma \rangle $ are mutually orthonormal still holds; using exact diagonalization in a $4 \times 4$-site system with periodic boundary conditions we verify that state overlaps remain $\ll 1$ except for Trugman loop configurations. In fact, for any state with strong local AFM correlations, we expect that this approximation is valid because the motion of the hole imprints a significant memory of its trajectory in the surrounding spin environment. This is found to be true even in a completely disordered spin environment at infinite temperature \cite{Kanasz-Nagy2017}, at least on a qualitative level. Because all typical states $| \Psi_{1/2} \rangle$ from the ensemble described by $\rho_{1/2}$ have significant local AFM order, we will assume in the following that the set of states $| \mathbf{j}, \sigma, \Sigma \rangle $ forms an orthonormal basis which defines the effective Hilbert space of the geometric string theory.
	
	Next we derive the effective Hamiltonian. For simplicity, we consider the $t-J$ Hamiltonian $\hat{\mathcal{H}}_{t-J} = \hat{\mathcal{H}}_t + \hat{\mathcal{H}}_J$ which provides an approximate low-energy description of the Fermi-Hubbard model when $U \gg t$. The first term, $\hat{\mathcal{H}}_t \propto t$, introduces couplings between string states $\langle \Sigma, \Sigma' \rangle$ corresponding to holes tunneling to neighboring sites on the Bethe lattice: $\hat{\mathcal{H}}^\Sigma_t = - t \sum_{ \langle \Sigma, \Sigma' \rangle } | \mathbf{j}, \sigma, \Sigma' \rangle \langle \mathbf{j}, \sigma, \Sigma | + {\rm h.c.}$. The spin-exchange part of the Hamiltonian, $\hat{\mathcal{H}}_J \propto J$, only depends on the spin configuration in the lattice. Because the strings distort this configuration, they can be associated with a finite potential energy $V_{\rm pot}(\Sigma) = \langle \mathbf{j}, \sigma, \Sigma | \hat{\mathcal{H}}_J | \mathbf{j}, \sigma, \Sigma \rangle$. In general, this expression depends on the specifics of the string configuration $\Sigma$. To simplify our model, we neglect self-interactions of the string and assume a linear string potential depending only on the string length $\ell_\Sigma$; thus in our effective model we consider the Hamiltonian $\hat{\mathcal{H}}^\Sigma_J = \sum_\Sigma V_{\rm pot}(\ell_\Sigma)$. The potential is derived by considering only straight strings, which yields $V_{\rm pot}(\ell_\Sigma) = (dE/d\ell) \ell_\Sigma + g_0 \delta_{\ell_\Sigma,0} + \mu_{\rm h}$; the linear string tension is $(dE/d\ell) = 2 J \left( C_s(\sqrt{2}) - C_s(1) \right)$ where $C_s(d)$ is the spin-spin correlator at distance $d$ as defined in the main text but for the undoped system, and the attractive potential $g_0 = -J \left( C_s(2) - C_s(1) \right)$ favors short strings. $\mu_{\rm h} = J (1 + C_s(2) - 5 C_s(1))$ denotes an overall energy offset which is irrelevant for our purposes. The most extreme self-interactions of the string, caused by loop configurations, are not expected to invalidate the geometric string approach; rather, they modify the hole dispersion and lead to additional dressing of the string with magnetic fluctuations \cite{Grusdt2018}. 
	
	Using the effective geometric string Hamiltonian $\hat{\mathcal{H}}^\Sigma = \hat{\mathcal{H}}^\Sigma_t + \hat{\mathcal{H}}^\Sigma_J$ introduced above, we can calculate the expected string length distribution. We consider a thermal state $\rho_\Sigma = e^{-\beta \hat{\mathcal{H}}^\Sigma} / Z_\Sigma$ for the string part. The overall state $\rho = \rho_{1/2} \otimes \rho_\Sigma$ factorizes and we use the experimental temperature $T=0.6 J$ throughout. This fixes the string tension $(dE/d\ell) = 0.85 J$, which we obtain by calculating the finite-temperature spin correlations $C_s(1), C_s(\sqrt{2})$ in the undoped Heisenberg model using a standard quantum Monte Carlo code from the ALPS package. We keep track of the exponentially large string Hilbert space by making use of the discrete rotational symmetries of the Bethe lattice which are present when the string potential depends only on the length $\ell_\Sigma$ of the string \cite{Grusdt2018}. The resulting string length distribution is shown in Fig.~1C of the main text; there, however, we show string lengths in units of sites rather than the bond count: the length of a string $\ell$ (in sites) is related to the length $\ell_\Sigma$ (in bonds) as $\ell = \ell_\Sigma + 1$.
	
	A few comments are in order. First, we fix the quantum numbers $\sigma$ and $\mathbf{j}$ specifying the beginning of the geometric string. However, spin-exchange processes introduce matrix elements between states with different initial positions, $| \mathbf{j}, \sigma, \Sigma \rangle $ and $| \mathbf{j}', \sigma, \Sigma' \rangle $, with a strength $ \propto J$ smaller than the dominant hopping amplitude $t > J$. As a result of such processes, we expect that $\mathbf{j}$ can be chosen randomly. Second, the beginnings of different fluctuating strings are expected to become correlated at sufficiently low temperatures. However, since their dynamics is determined by an energy scale $J$, and the experimental temperature is of similar order of magnitude, we expect that such correlations between $\mathbf{j}_1$ and $\mathbf{j}_2$ associated with two different holes can be neglected in the current experimental regime. Third, thermal excitations of the fluctuating strings include vibrational and rotational \cite{Grusdt2018} excitations. If rotational excitations are ignored, a significantly narrower string length distribution is obtained which is dominated by quantum fluctuations. Indeed, at somewhat higher temperatures $T \approx 0.8 J$, we find from our effective model that the string length diverges because the free energy can be reduced by creating a high-entropy state with exponentially many rotational excitations. This transition is predicted in a regime where the experimental sample is too hot to measure a string signal which differs significantly from an infinite temperature state. 
	
	Our approach is based on earlier work by Bulaevskii et al. \cite{Bulaevskii1968} and later by Brinkman and Rice \cite{Brinkman1970} and Trugman \cite{Trugman1988}, where similar calculations with strings have been performed at zero temperature and considering a classical N\'eel state. The frozen-spin approximation represents an approximate way of generalizing these results to situations with quantum and thermal fluctuations. The obtained trial wavefunction can also be interpreted as a microscopic formulation of the meson picture of magnetic polarons: instead of the most common description of holes as heavily dressed by magnetic fluctuations \cite{Schmitt-Rink1988,Kane1989,Sachdev1989,Martinez1991,Liu1992}, this theory -- originally proposed by B\'eran et al. \cite{Beran1996} using phenomenological arguments -- describes the doped holes as bound states of spin-less chargons and charge-neutral spinons. Including the properties of the spinon, located at the opposite end of the geometric string from the chargon, is essential for recovering the known microscopic properties of a single hole in an AFM. On a macroscopic level, the geometric-string theory discussed here describes a fermionic gas of mesons -- a candidate state which has also been proposed for the elusive pseudogap phase in cuprates \cite{Baskaran2007,Chowdhury2015,Punk2015}.

	\subsection{$\pi$-flux theory}
	
	We use Metropolis Monte Carlo sampling to obtain Fock states of fermions described by the Gutzwiller projected thermal density matrix $\hat{\rho} = \mathcal{P}_{GW} e^{-\hat{H}_{MF}/k_BT}\mathcal{P}_{GW}$ determined by the quadratic Hamiltonian
	\begin{equation}
	\begin{split}
	\hat{H}_{MF} = -\frac{1}{2} J^* \sum_{\mathbf{i} \in A} \sum_\sigma \left( e^{i\theta_0} \hat{c}_{\mathbf{i},\sigma}^\dagger \hat{c}_{\mathbf{i}+\mathbf{x},\sigma} + e^{-i\theta_0} \hat{c}_{\mathbf{i},\sigma}^\dagger \hat{c}_{\mathbf{i}+\mathbf{y},\sigma} + h.c. \right) 
	\\
	-\frac{1}{2} J^* \sum_{\mathbf{i} \in B} \sum_\sigma \left( e^{-i\theta_0} \hat{c}_{\mathbf{i},\sigma}^\dagger \hat{c}_{\mathbf{i}+\mathbf{x},\sigma} + e^{i\theta_0} \hat{c}_{\mathbf{i},\sigma}^\dagger \hat{c}_{\mathbf{i}+\mathbf{y},\sigma} + h.c. \right).
	\end{split}
	\end{equation}
	Here, $\mathbf{i} \in A(B)$ denotes lattice sites $\mathbf{i}$ which are part of the A(B) sublattice and $\hat{c}_{\mathbf{i},\sigma}^{(\dagger)}$ is the annihilation (creation) operator of a fermion with spin $\sigma$. The Hamiltonian describes a system with staggered flux $\pm \Phi = \pm 4 \theta_0$, and the $\pi$-flux state is characterized by $\theta_0 = \frac{\pi}{4}$ \cite{Marston1989}. Fourier transforming leads to
	\begin{equation}
	\hat{H}_{MF} = \sum_{\mathbf{k} \in MBZ,\sigma} \begin{pmatrix} \hat{c}_{\mathbf{k},\sigma}^\dagger & \hat{c}_{\mathbf{k}+\mathbf{G},\sigma}^\dagger \end{pmatrix} \hat{h}_{\mathbf{k}} \begin{pmatrix}
	\hat{c}_{\mathbf{k},\sigma} \\ \hat{c}_{\mathbf{k}+\mathbf{G},\sigma}
	\end{pmatrix}
	\end{equation}
	with 
	\begin{equation}
	\hat{h}_{\mathbf{k}} =  \text{Re}{R(\mathbf{k})} \hat{\tau}^z + \text{Im}{R(\mathbf{k})} \hat{\tau}^y
	\end{equation}
	and
	\begin{equation}
	R(\mathbf{k}) = -J^* \left(\cos k_x e^{i\theta_0} + \cos k_y e^{-i \theta_0}\right),
	\end{equation}
	with Pauli matrices $\hat{\mathbf{\tau}} = \left(\hat{\tau}^x,\hat{\tau}^y,\hat{\tau}^z\right)$.
	Here, $\mathbf{k}\in MBZ$ denotes momenta $\mathbf{k}=\left( k_x, k_y\right)$ in the magnetic Brillouin zone. Diagonalizing $\hat{h}_{\mathbf{k}}$ leads to two eigenstates $\left| u_{\mathbf{k},\mu} \right> = \left( u_{\mathbf{k},\mu}^0, u_{\mathbf{k},\mu}^1 \right)^T $ for every momentum $\mathbf{k}$ in the magnetic Brillouin zone. Bloch's theorem yields the wave function
	\begin{equation}
	\psi_{\mathbf{k},\mu}(\mathbf{r}) = \frac{1}{\sqrt{L^2}} \left( u_{\mathbf{k},\mu}^0 + u_{\mathbf{k},\mu}^1 e^{-i \mathbf{G}\mathbf{r}} \right) e^{-i\mathbf{k}\mathbf{r}}
	\end{equation}
	with band index $\mu = \pm$ and an $L \times L$-site system size. We consider a system of 16 by 16 sites and cut out a circular region of interest of the same size as in the experiment to obtain the same boundary effects in both cases. For a given doping value, we assume a spin balanced system and start from a random occupation of states in momentum space for both up and down fermions as well as a random configuration without double occupancies in real space. The exclusion of states with double occupancies in the sampling corresponds to applying the Gutzwiller projection. From any given state, updates in real space as well as updates in momentum space for up and down fermions are possible. In real space, two neighboring sites can exchange their occupation if they differ. In momentum space, a given fermion can change its momentum to any other unoccupied momentum. Note that in momentum space, we treat up and down fermions separately from each other such that two fermions of opposite spin can have the same momentum. 
	The snapshots are generated by Metropolis Monte Carlo sampling according to the probability distribution
	\begin{equation}
	p_{\beta}(\alpha_{\mathbf{r}},\alpha_{\mathbf{k}}) = e^{-\beta E(\alpha_{\mathbf{k}})} |\left<\alpha_{\mathbf{r}} | \alpha_{\mathbf{k}} \right> |^2,
	\label{eq:probDistr}
	\end{equation}
	where $\beta = \left(k_B T \right)^{-1}$ is the inverse temperature and $\left| \alpha_{\mathbf{r} (\mathbf{k})} \right>$ denote Fock states in configuration (momentum) space. Note that Eq.~\eqref{eq:probDistr} is not normalized. However, the normalization does not matter for the Metropolis sampling, since only ratios of probability distributions are required.
	
	The energy of a state $\left| \alpha_{\mathbf{k}} \right>$ is given by
	\begin{equation}
	E(\alpha_{\mathbf{k}}) = \sum_{\mathbf{k} \text{ occ. in }\alpha_{\mathbf{k}}} \epsilon(\mathbf{k})
	\end{equation}
	with eigenenergies $\epsilon(\mathbf{k})$ of $\hat{h}_{\mathbf{k}}$ and the sum is taken over momenta $\mathbf{k}$ which are occupied in the considered Fock state $\left| \alpha_{\mathbf{k}} \right>$.  
	
	After generating a sample of several thousand Fock states $\alpha_r$, doublon-holon pairs are artificially added with a probability given by $4 t^2/U^2$ on nearest-neighbor sites with opposite spins. The experimentally measured anti-moment correlator as shown in Fig.~5 of the main text, as well as numerics \cite{Cheuk2016a}, indicate that restricting doublon-hole pairs to nearest neighbors is a valid approximation in this regime. In the following, we consider a region of interest of the same size and shape as in the experiment. Furthermore, we simulate the experimental imaging procedure and keep the parity-projected density distribution of either both spins or with one spin state removed. The coupling $J^*=3J$ in the mean field Hamiltonian is chosen such that at half filling, the nearest- and next-nearest-neighbor spin correlators obtained from the simulation at the temperature $T = 0.6J$ fit the experimental data as closely as possible. Without any other fitting parameter, the doping dependence of the nearest neighbor spin correlator is described correctly. However, the temperature dependence of the spin correlators even at half filling is not captured correctly with these parameters. 
	
	\subsection{Free chargon theory}
	For reference, we consider a purely phenomenological theory of free fermionic chargons in the intermediate doping regime above $\delta > 5 \%$. Theoretically, it is motivated by the possibility that spinon-chargon pairs unbind and a deconfined phase of chargons may be realized. For simplicity we consider free fermions, although qualitatively similar anti-correlations would be expected for a gas of bosonic chargons with hard core repulsion. More informed theoretical work has also proposed the possibility of a non-trivial metallic state of chargons \cite{Kaul2007}. In our present work, we compare to a free chargon theory to calculate the hole (or anti-moment) correlations. The phenomenological model assumes point-like fermionic chargons $\hat{h}_{\mathbf{j}}$ on the square lattice, with an effective Hamiltonian $\hat{\mathcal{H}}_{\rm ch} = - t \sum_{\langle \mathbf{i}, \mathbf{j} \rangle} \hat{h}^\dagger_{\mathbf{j}} \hat{h}_{\mathbf{i}} + {\rm h.c.}$ and with the largest conceivable hopping strength $t$ between neighboring sites. The chargon-chargon correlations are then calculated from a simple thermal state $\rho_{\rm ch} = e^{- \beta \hat{\mathcal{H}}_{\rm ch}}$ with $\beta = 1 / k_{\rm B} T$ for the experimental temperature $T=0.6 J$ and $t=2 J$, see Fig. 5 in the main text. 
	
	\subsection{Point-like magnetic polaron theory}
	We compare the experimentally measured anti-moment correlations to a model of free point-like magnetic polarons with the known dispersion of a free hole in an AFM \cite{Martinez1991,Liu1992}. To this end we consider a model of free, point-like, fermionic magnetic polarons $\hat{m}_{\mathbf{j}}$ on the square lattice, with a momentum-space Hamiltonian $\hat{\mathcal{H}}_{\rm mp} = \sum_{\mathbf{k}} \hat{m}^\dagger_{\mathbf{k}} \hat{h}_{\mathbf{k}} \epsilon_{\rm mp}(\mathbf{k})$. The dispersion relation was approximated as:
	\begin{equation}
	\epsilon_{\rm mp}(\mathbf{k}) = \left[ 4 \chi^2 J^2 | \cos(k_x) e^{- i \Phi/4} + \cos(k_y) e^{i \Phi/4} |^2 + B_{\rm st}^2 / 4 \right]^{1/2},
	\end{equation}
	motivated by the mean-field description of the staggered flux plus N\'eel state \cite{Lee1988} with parameters $\chi=0.8$, $B_{\rm st} = 0.35 J$ and $\Phi=0.4 \pi$ determined such that the exact quantum Monte Carlo results at $J=0.4 t$ \cite{Brunner2000} are correctly captured. The resulting two-point correlations for a range $T=0.5 J$ to $T=0.7 J$ in the thermal state $\rho_{\rm mp} = e^{- \beta \hat{\mathcal{H}}_{\rm mp}} / Z_{\rm mp}$ are calculated for $t=2 J$ in the main text.

	\section{Correlation functions}
	
	\subsection{Temperature dependence of $C_s(d)$}
	The data presented in Fig.~4 of the main text includes samples with temperatures between $0.5J$ and $0.7J$, binned by doping values with $2\%$ resolution. In Fig.~S9A we plot $C_s(1)$, $C_s(\sqrt{2})$, and $C_s(2)$ versus doping for each individual experimental dataset, where colorbars for each quantity denote temperature. While it is clear that colder temperatures are accompanied by stronger correlations, crucially one can see that the zero crossing of $C_s(\sqrt{2})$ persists across the entire temperature range included.
	
	We can also compare the experimentally measured $C_s(1)$ versus doping to determinantal quantum Monte Carlo calculation of the Hubbard model on an $8 \times 8$ homogeneous square lattice using the Quantum Electron Simulation Toolbox, see Fig.~S9B \cite{Varney2009, Brown2019}. Agreement between the two indicate that our experimental approach to doping the system does not change the temperature of the sample beyond experimental uncertainty.
	
	\subsection{Doping dependence of larger distance correlations}
	As an extension of Fig.~4A of the main text, in Fig.~S10 we plot $C_s(d)$ for $d = \sqrt{2}$, $2$, $\sqrt{8}$, and $3$ for experiment, geometric strings, and $\pi$-flux states. Statistical uncertainty makes it challenging to quantitatively compare experiment with theory. However, examining all cases independently it appears that for all of them, larger-distance correlators may not exhibit a sign change with doping. In the case of geometric strings, the mixing of correlators beyond the nearest neighbor correlator demonstrates a much smaller effect than for the nearest neighbor correlator because longer-distance correlation lengths are much more similar in magnitude.
	
	\newpage
	
	\tocless{\section*{Fig S1}}
	
	\begin{figure}[h]
		\centering
		\includegraphics[width=\textwidth]{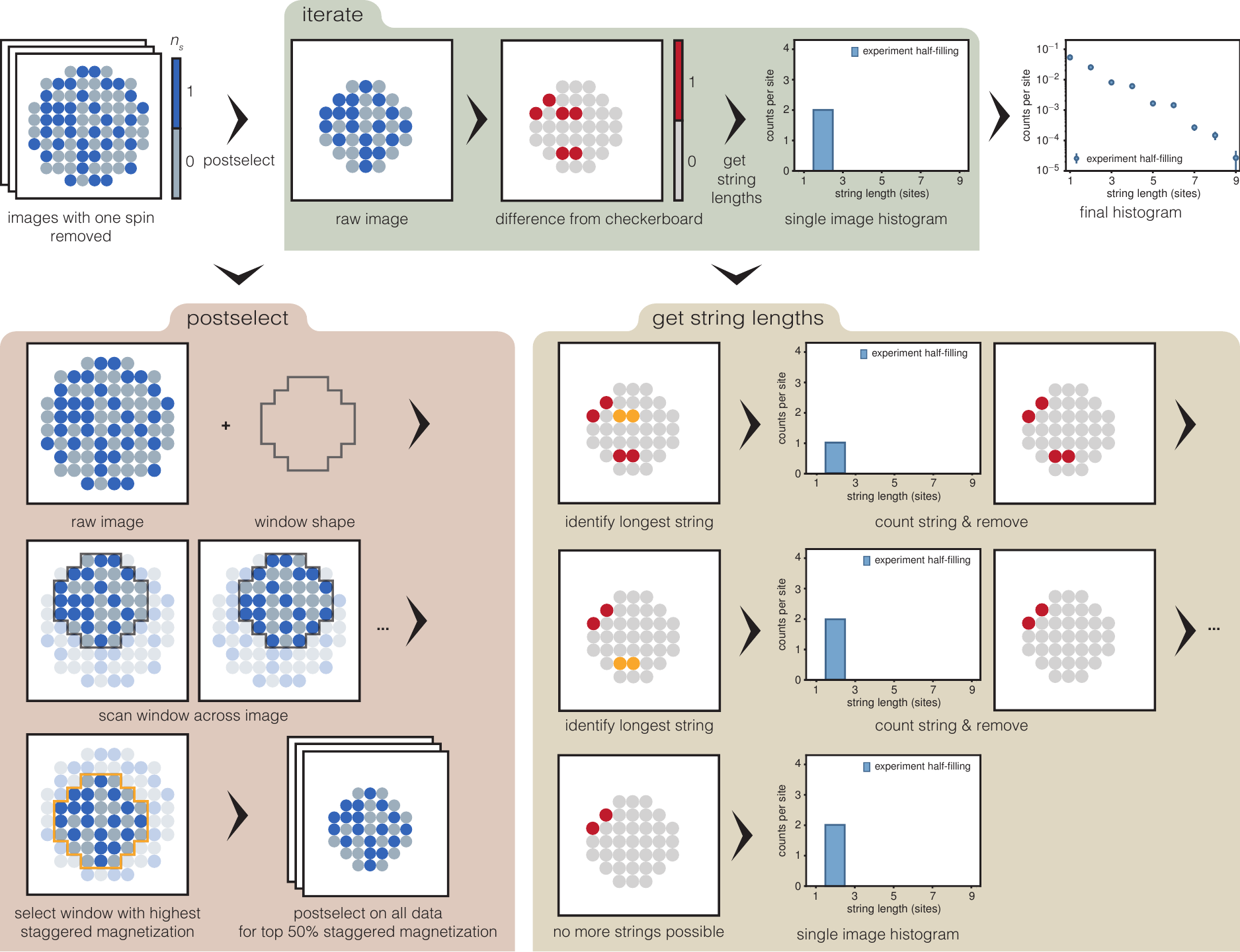}
		\caption{String detection algorithm. This schematic outlines the string detection algorithm used in the main text and detailed in section 3.1 of the supplementary text. The algorithm only uses images with one spin species removed. After postselection, the deviation from a reference checkerboard pattern is used to identify string patterns.}
	\end{figure}
	
	\clearpage
	
	\tocless{\section*{Fig S2}}
	
	\begin{figure}[h]
		\centering
		\includegraphics{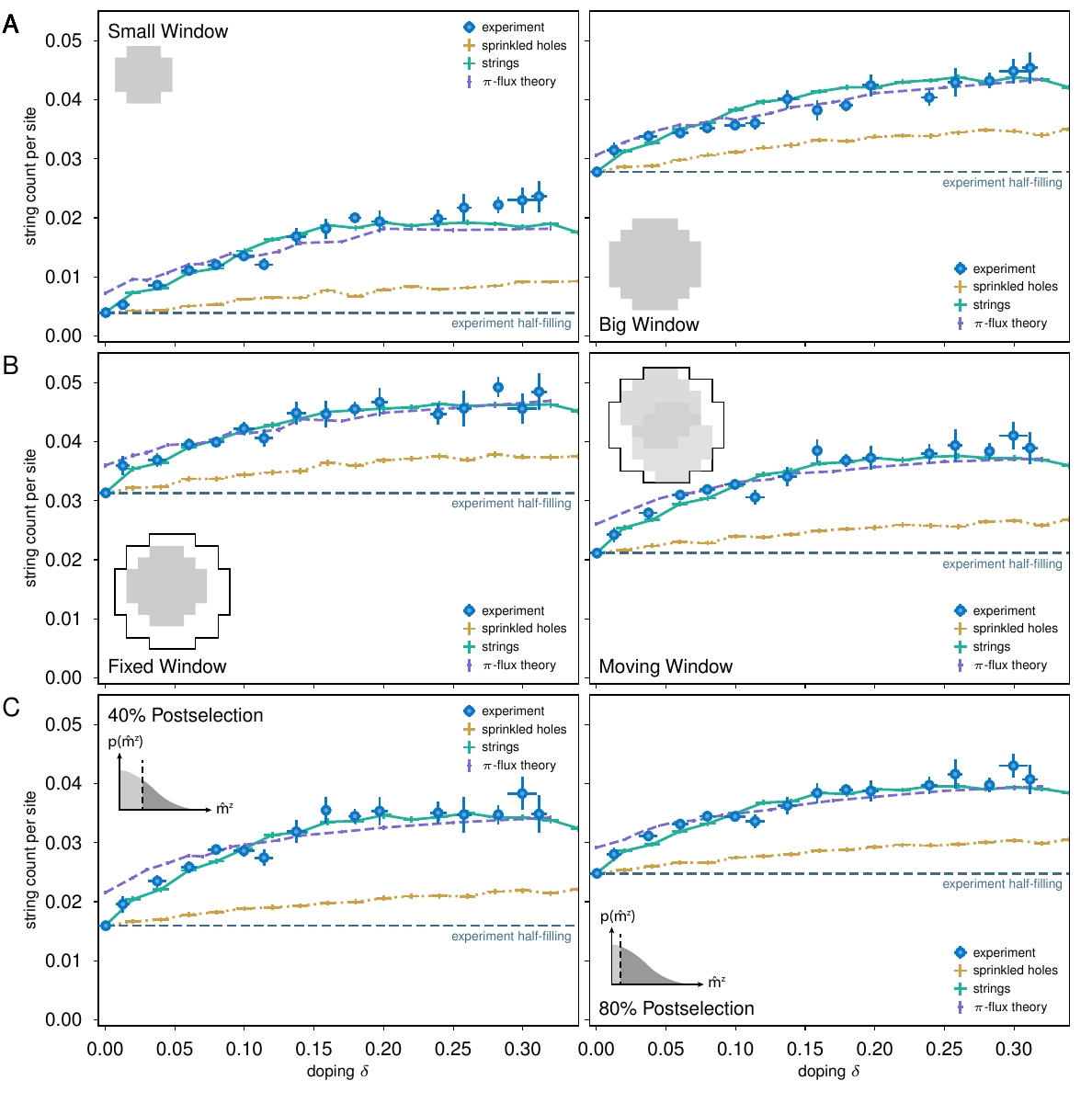}
		\caption{Effect of string count on post-selection. All parameters not mentioned are kept fixed as in the procedure outlined in the main text. In all cases, we see that the qualitative features described in the main text are maintained. (A) Using a different-size window for the analysis region, either 5 or 8 sites in diameter. (B) Fixing the window position to the center of the system, compared to scanning the window position to maximize the staggered magnetization. (C) Varying the percentage of data kept when postselecting on the staggered magnetization, either $40\%$ or $80\%$.}
	\end{figure}
	
	\clearpage
	
	\tocless{\section*{Fig S3}}
	
	\begin{figure}[h]
		\centering
		\includegraphics{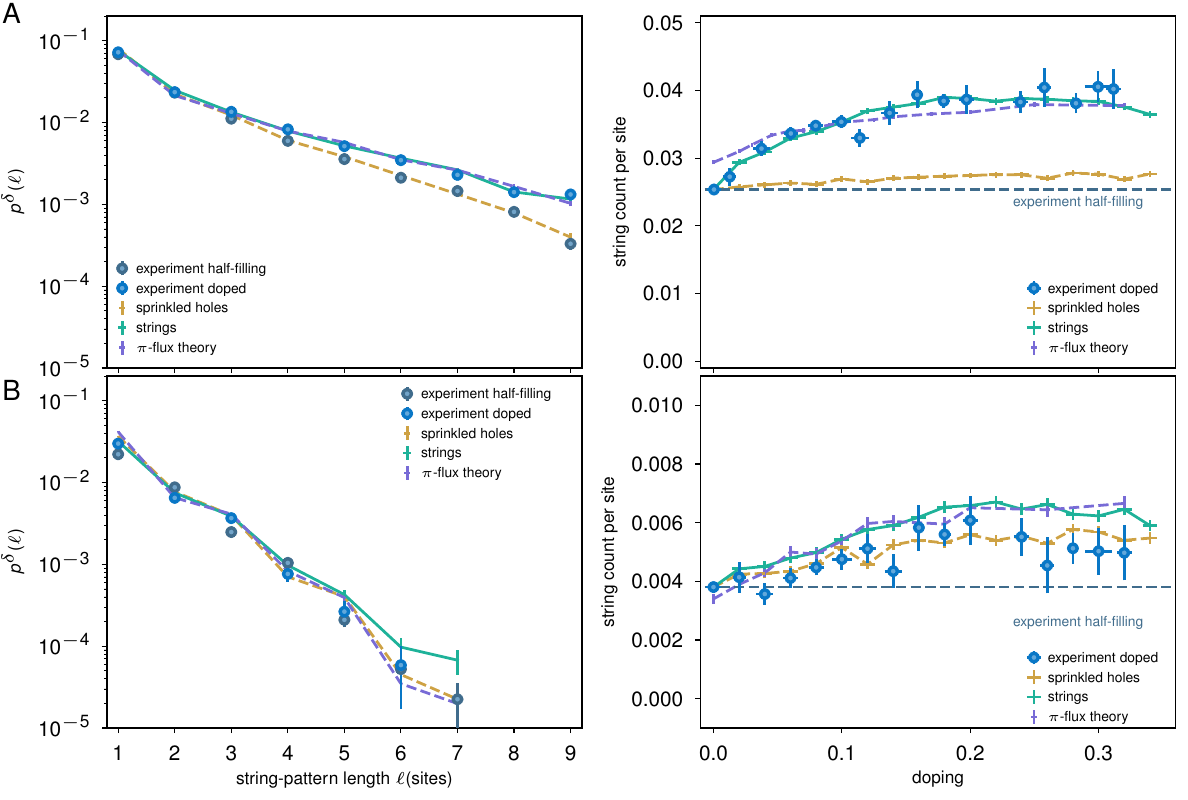}
		\caption{Alternate string detection schemes. (A) The string length histogram and total string count produced with the simplified difference pattern extraction algorithm. Apart from an increase in values of the string-pattern-based observables at half-filling, the results are similar to the detection algorithm used in the main text. (B) Same quantities, but produced with the happiness method, are also sensitive to doping but may be too noisy to be useful (see supplementary text section 3.4).}
	\end{figure}
	
	\clearpage
	
	\tocless{\section*{Fig S4}}
	
	\begin{figure}[h]
		\centering
		\includegraphics{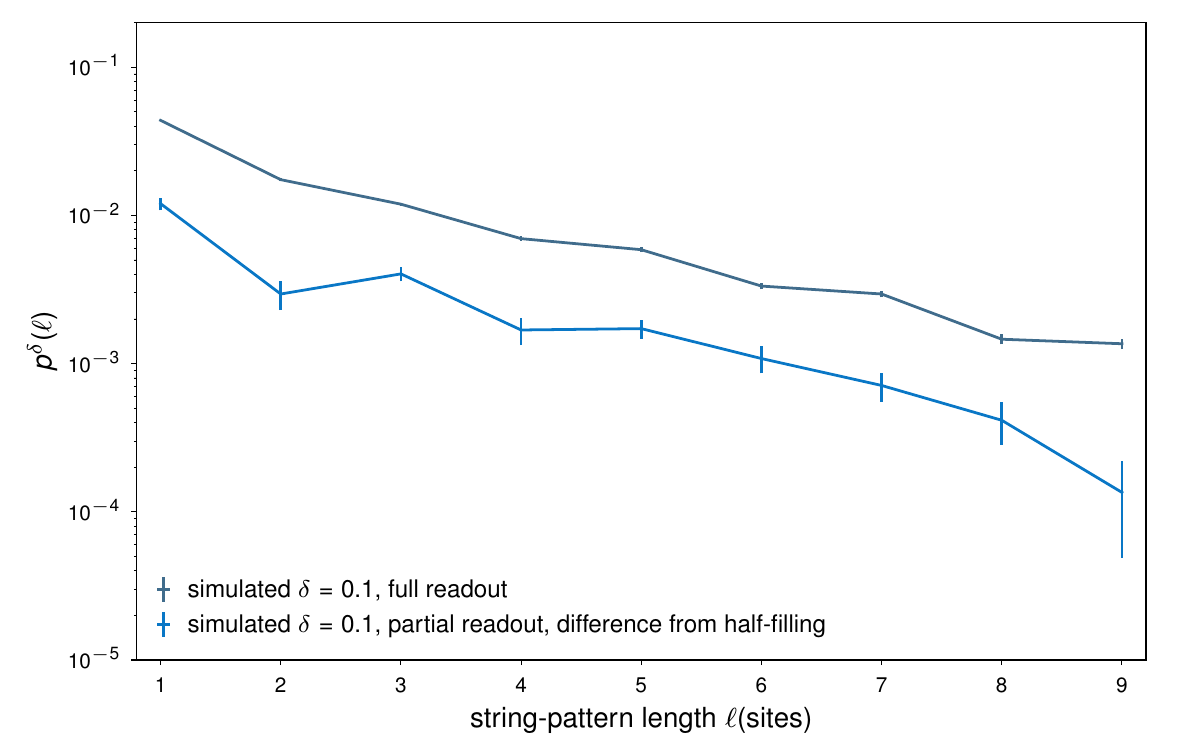}
		\caption{String pattern detection with simulated full spin readout and partial readout (with half-filling signal subtracted) using Heisenberg QMC data with added charge fluctuations and strings. The signals have qualitatively the same shape but are offset due to the lower detection efficiency without full spin readout.}
	\end{figure}
	
	\clearpage
	
	\tocless{\section*{Fig S5}}
	
	\begin{figure}[h]
		\centering
		\includegraphics{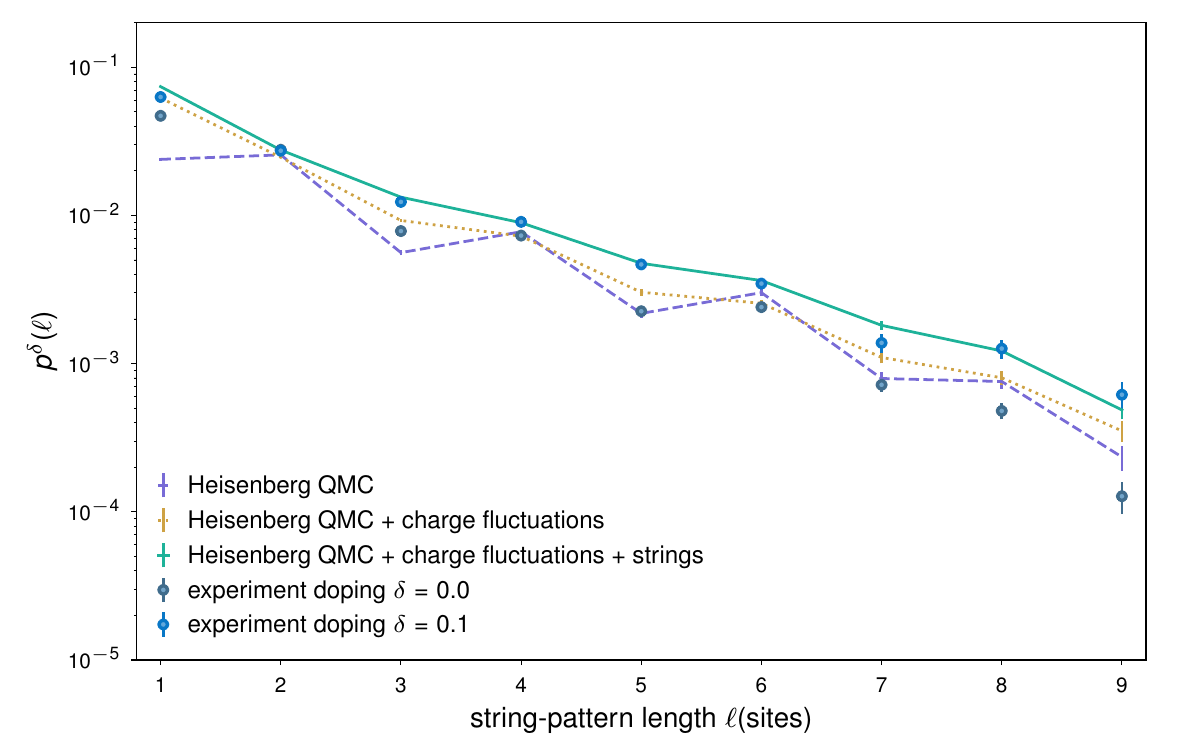}
		\caption{Measured string pattern distribution for Heisenberg QMC simulation at half-filling and for $10\%$ doping. Without introducing charge fluctuations in the Heisenberg simulation, we see a significant deviation from experiment in the half-filling distribution at low lengths. However, the simulation including charge fluctuations qualitatively matches experimental data, providing evidence that the string patterns detected in the experimental data at half-filling arise from the distinction between snapshots of a quantum antiferromagnet and the reference checkerboard.}
	\end{figure}
	
	\clearpage
	
	\tocless{\section*{Fig S6}}
	
	\begin{figure}[h]
		\centering
		\includegraphics{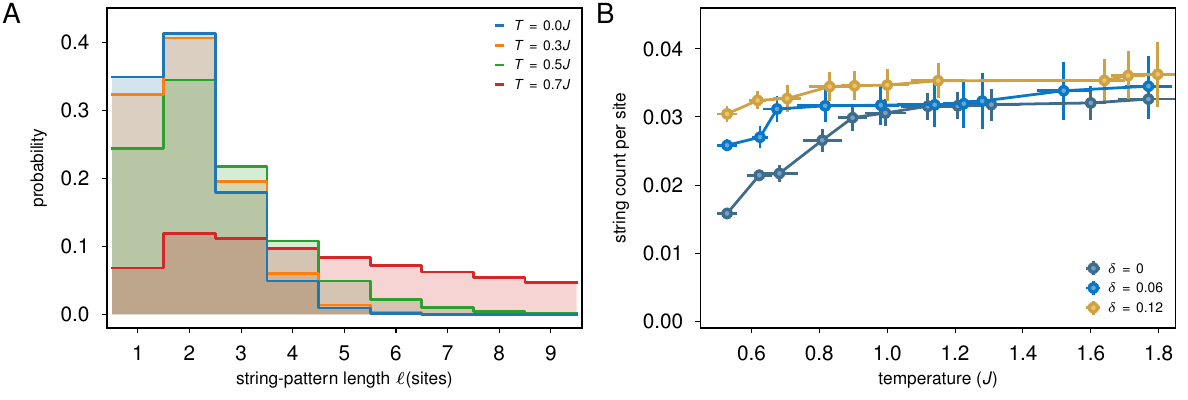}
		\caption{Effects of temperature on string states and detection. (A) Analytic string length distribution for various temperatures. Close to around $T=0.8J$, the strings become unbound and the histogram significantly broadens. (B) Effect of increased temperature on total string count, for various doping values. Saturation occurs at a lower temperature for higher doping, reflecting how the string count can increase with temperature and doping. In particular, increasing temperature destroys the AFM spin background, but under the string picture it also increases the length of geometric strings. Under our detection scheme, the former dominates, and we cannot isolate the latter effect.}
	\end{figure}
	
	\clearpage
	
	\tocless{\section*{Fig S7}}
	
	\begin{figure}[h]
		\centering
		\includegraphics[width=\textwidth]{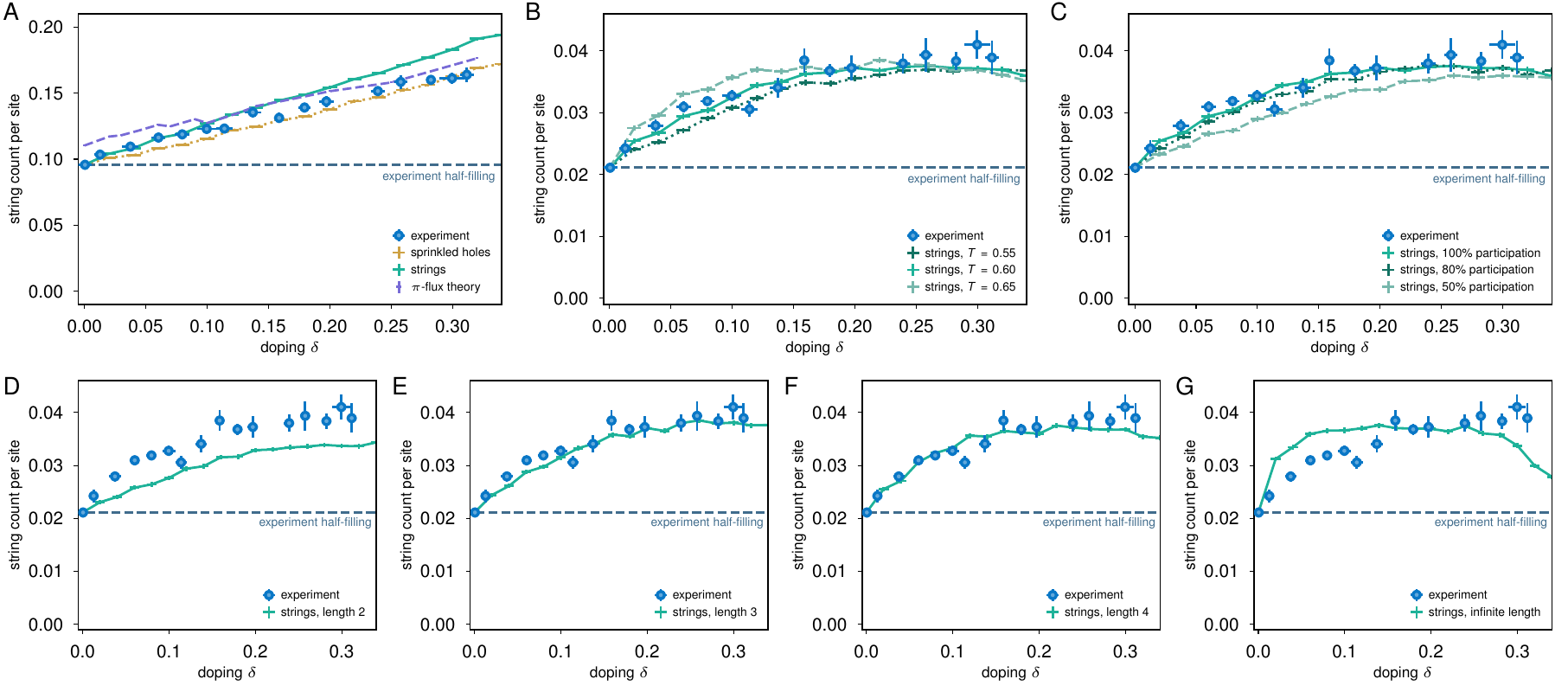}
		\caption{Effect of analytic string length distribution on measured string count. (A) String count versus doping, as in Fig.~3A of the main text, but including all string-pattern lengths in the count. The agreement between all simulations and experiment suggests that this quantity may be trivially dependent on doping. (B) Comparison of experimentally measured string count and simulated geometric strings, for analytic string length distributions corresponding to temperatures $T/J$ of 0.55, 0.60 (as in main text), and 0.65. The $T/J=0.60$ distribution matches experiment best. (C) Same, but varying the fraction of simulated holes which are then propagated to simulate strings. The best agreement occurs when all holes are part of strings. (D-G) Same, but for simulated strings which are of all of a single length, ranging from 2 sites to the infinite-site limit, instead of sampled from the analytic distribution as described in the main text. In (B-G), the experimental result shown is the same.}
	\end{figure}
	
	\clearpage
	
	\tocless{\section*{Fig S8}}
	
	\begin{figure}[h]
		\centering
		\includegraphics[width=\textwidth]{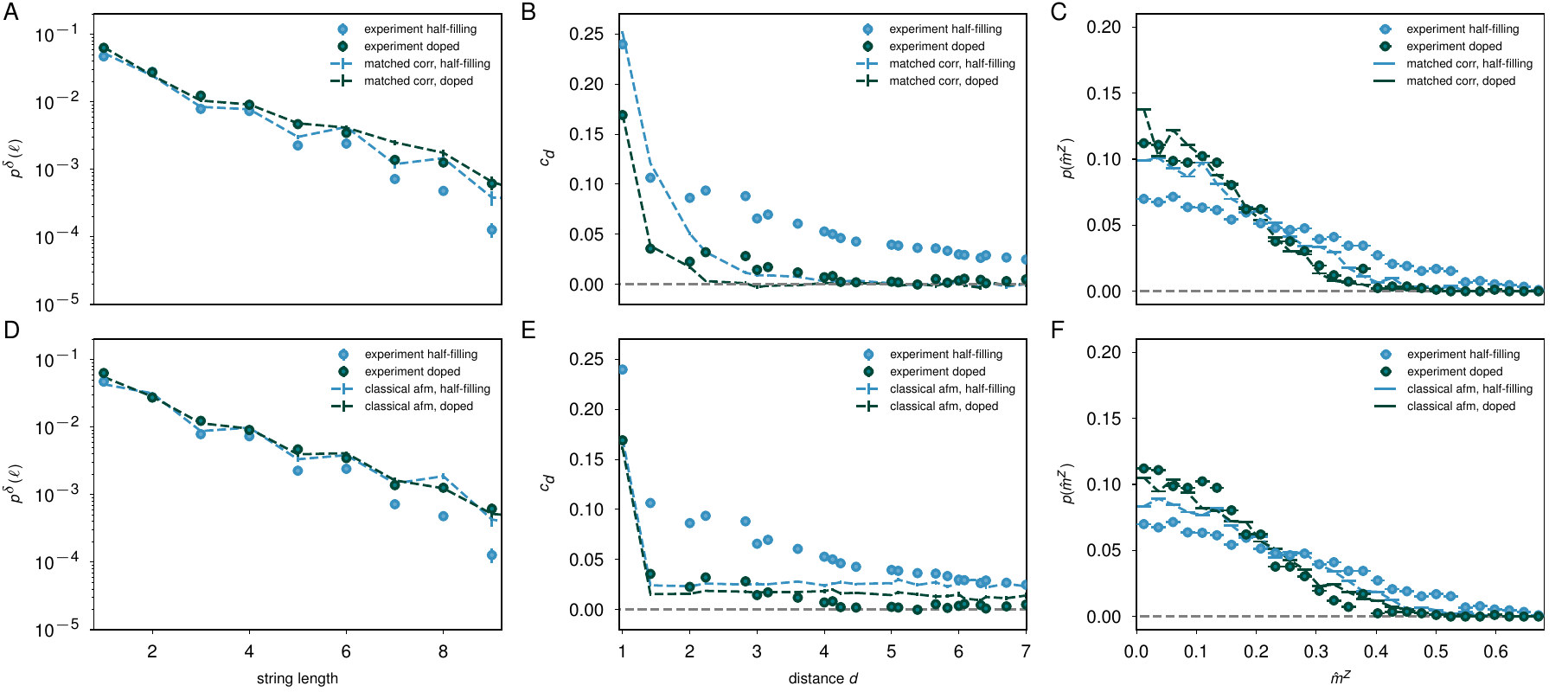}
		\caption{Performance of phenomenological models. (A-C) String length histogram, spin correlation function, and staggered magnetization distribution for experiment and a phenomenological model where spins are flipped in a random spin distribution until the nearest-neighbor and next-nearest neighbor spin correlators match experimental values. (D-F) Same quantities, but comparing experiment to a phenomenological model where singlet pairs are added to a classical N\'eel checkerboard and a projective measurement is performed. The singlet pair density is tuned to achieve rough agreement with the doped experimental data.}
	\end{figure}
	
	\clearpage
	
	\tocless{\section*{Fig S9}}
	
	\begin{figure}[h]
		\centering
		\includegraphics[width=\textwidth]{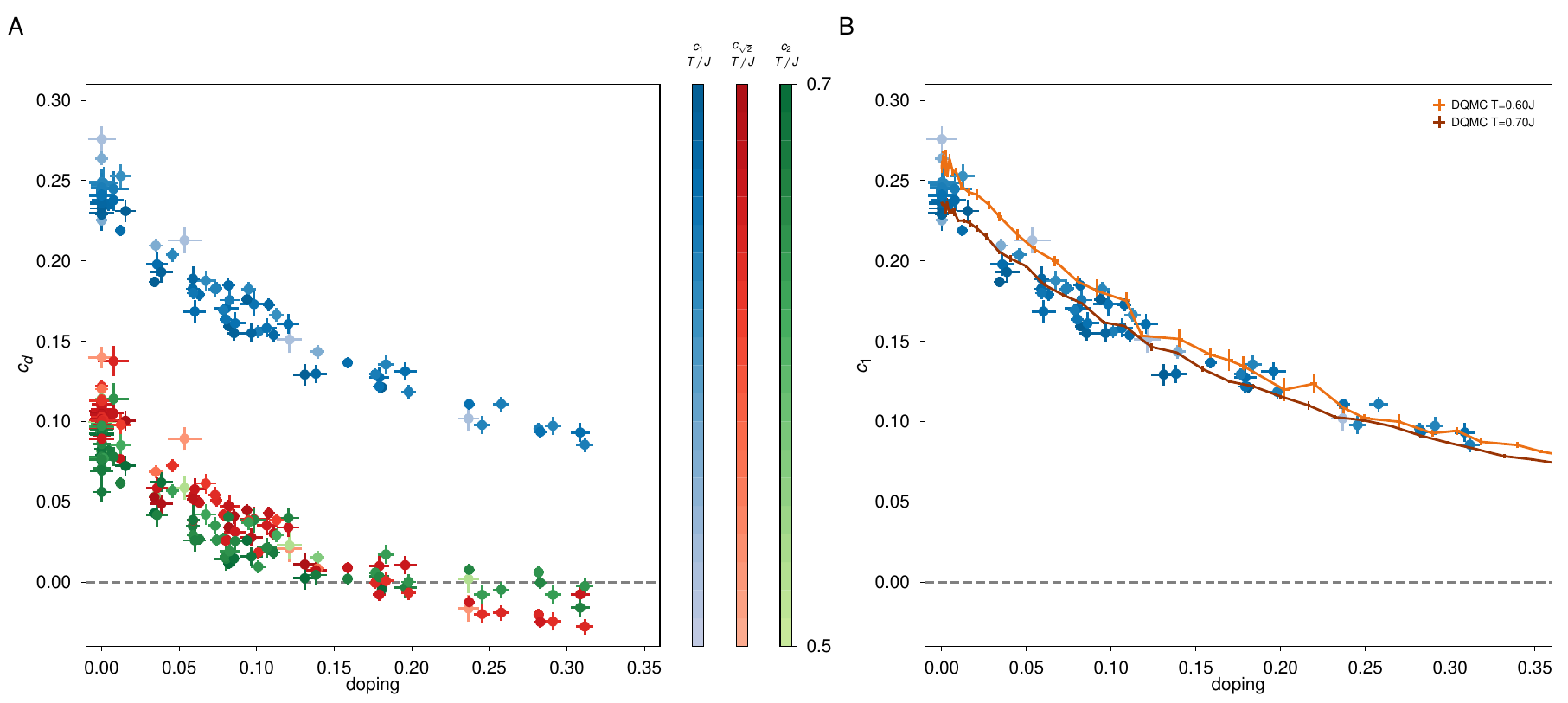}
		\caption{Temperature and spin-spin correlations. (A) Replotting the data from Fig.~4 of the main text prior to grouping by doping shows additional spread from statistical fluctuation and temperature. The temperature of each dataset is indicated by the colorbars; the average temperature of all datasets, weighted by dataset size, is $T = 0.65(4)J$. (B) Comparison of the measured nearest-neighbor spin correlator to determinantal quantum Monte Carlo simulation shows that as we dope the system, its temperature does not increase.}
	\end{figure}
	
	\clearpage
	
	\tocless{\section*{Fig S10}}
	
	\begin{figure}[h]
		\centering
		\includegraphics[width=\textwidth]{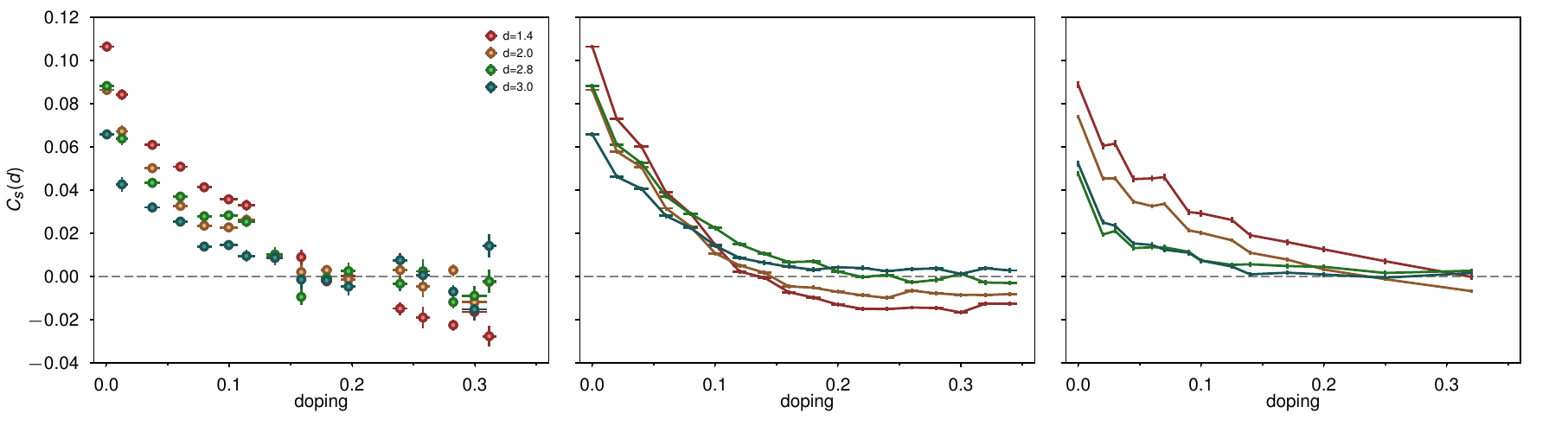}
		\caption{Larger distance spin-spin correlators $C_s(d)$ from experiment at $T = 0.65(4)J$ (left), geometric-string theory (middle), and $\pi$-flux states (right).}
	\end{figure}
	
	\clearpage
	\nocite{*}
	\tocless{}
	
\end{document}